\documentclass[a4paper,usenatbib]{mnras}

\makeatletter

\newcommand{\Rmnum}[1]{\expandafter\@slowromancap\romannumeral #1@}

\newcommand{\gsim}{\lower0.6ex\vbox{\hbox{$\buildrel{\textstyle >}\over{\sim}\ $}}}

\makeatother

\usepackage{newtxtext,newtxmath}
\usepackage{graphicx}	
\usepackage{amsmath}	
\usepackage{subcaption}
\usepackage{booktabs}
\usepackage[dvipsnames,svgnames]{xcolor}
\usepackage{color}

\newcommand{\yueying}[1]{\textcolor{black}{#1}}
\newcommand{\yy}[1]{\textcolor{black}{#1}}
\newcommand{\yn}[1]{\textcolor{black}{#1}}

\def\hmpc{h^{-1}{\rm Mpc}}

\def\invhmpc{\;h\;{\rm Mpc}^{-1}}

\def\hkpc{h^{-1}\, {\rm kpc}}
\def\hckpc{h^{-1}\, {\rm ckpc}}
\def\kms{\, {\rm km}\, {\rm s}^{-1}}

\def\kmsmpc{\kms\;{\rm Mpc}^{-1}}

\title[BH-growth in Constrained Simulations]{Not all peaks are created equal: the early growth of Supermassive Black Holes}

\author[Y. Ni et al.]
{Yueying Ni$^{1}$\thanks{Email:yueyingn@andrew.cmu.edu}, Tiziana Di Matteo$^{1}$, Yu Feng$^{2}$ \\
$^1$ McWilliams Center for Cosmology, Department of Physics, Carnegie Mellon University, Pittsburgh, PA 15213 \\
$^2$ Berkeley Center for Cosmological Physics and Department of Physics, University of California, Berkeley, CA 94720, USA \\
}

\date{Accepted XXX. Received YYY; in original form ZZZ}

\pubyear{2020}
\begin{document}
\maketitle

\begin{abstract}
In this work, we use the constrained Gaussian realization technique to study the early growth of supermassive black holes (SMBHs) in cosmological hydrodynamic simulations, exploring its relationship with features of the initial density peaks on large scales, $\sim$ 1 $\hmpc$.
Our constrained simulations of volume (20 $\hmpc$)$^3$ successfully reconstruct the large-scale structure as well as the black hole growth for the hosts of the rare $10^9$ $M_{\odot}$ SMBHs found in the \textsc{BlueTides} simulation at $z \sim 7$.
We run a set of simulations with constrained initial conditions by imposing a $5 \sigma_0(\rm R_G)$ peak on the scale of $\rm R_G = 1$ $\hmpc$ varying different peak features, such as the shape and compactness as well as the tidal field surrounding the peak.
We find that initial density peaks with high compactness and low tidal field induce the most rapid BH growth at early epochs. This is because compact density peaks with a more spherical large scale matter distribution lead to the formation of the highest gas inflows (mostly radial) in the centers of halos which boost the early BH accretion.
Moreover, such initially compact density peaks in low tidal field regions also lead to a more compact BH host galaxy morphology. Our findings can help explain the tight correlation between BH growth and host galaxy compactness seen in observations.
\end{abstract}


\section{Introduction}
\label{section1:introduction}

Recently, there has been much ongoing progress in the detection of $z>6-7$ luminous quasi-stellar objects (QSOs): more than 200 QSOs have been discovered beyond $z=6$ \citep[see,e.g.][and references therein]{Fan2019}, and a handful found with $z>7$ \citep{Mortlock2011, Wang2018, Banados2018, Matsuoka2019, Yang2019, Wang2021}.
These QSOs must be powered by the first supermassive black holes (SMBHs) with masses $M_{\rm BH} \sim 10^9 M_{\odot}$ that grow within the first billion years of the Universe. The existence and formation of these giant BH "monsters" in the early universe remain one of the greatest challenges in our standard paradigm of structure formation.

Large cosmological simulations are now a useful tool to study the formation and evolution of SMBHs in the high redshift universe ($z > 6$).
Since the bright QSOs are extremely rare, simulations need to cover a huge volume to probe these highly biased environments to compare directly with the currently observed high-z quasar samples.
The \textsc{BlueTides} is a large volume (400 $\hmpc$ per side), high resolution (with $2 \times 7040^3$ particles) cosmological hydrodynamic simulation that has run to $z = 6.5$. Its large volume and high resolution provide an ideal suite to investigate the first generation of massive galaxies and the rare QSOs in the high-$z$ universe.
So far, the galaxy formation model of \textsc{BlueTides} has been tested against various observations of the high-$z$ universe and is in good agreement with all current observational constraints, such as the UV luminosity functions \citep{Feng2016, Waters2016a, Waters2016b, Wilkins2017}, the properties of the first galaxies and the most massive quasars \citep{Feng2015, DiMatteo2017, Tenneti2018}, the galaxy stellar mass functions \citep{Wilkins2018}, the angular clustering of galaxies \citep{Bhowmick2017}, the BH-galaxy scaling relations \citep{Huang2018}, as well as the gas outflows from the $z=7.54$ quasar \citep{Ni2018}, the BH host galaxies \citep{Marshall2019} and the high-$z$ obscured AGN population \citep{Ni2020}.

Early QSOs are extremely rare implying that not all massive halos will host a supermassive BH at high redshifts.
For example, in the (400 $\hmpc$)$^3$ volume of the \textsc{BlueTides} simulation, there is only one SMBH with mass above $10^8 M_{\odot}$ at $z=8$ residing in a halo host with a mass of $\sim 10^{12} M_{\odot}$, while there are > 50 halos more massive than that SMBH host halo.
Therefore it is interesting to study whether and how the early growth and evolution of SMBHs is related to their surrounding large-scale matter distribution and environment.
For example, using the \textsc{BlueTides} simulation \cite{DiMatteo2017} found that, the most massive BHs reside in the most isolated high-density peak characterized by a low tidal field strength on scales of $\sim 1 \hmpc$. 
The correlation between tidal field strength and BH growth in the centre of the halo appeared to be a key factor contributing to the formation of the first quasars.
In this work, we apply the Constrained Realization (CR) technique to explicitly impose constraints on the large-scale features of the primordial density field in the initial conditions (IC) to test and verify these findings.
We carry out a series of hydrodynamic simulations using constrained ICs with various sets of constraint parameters (including peak height, compactness, tidal field, etc.) to explore the effects of the large-scale environment on the early BH growth in highly biased regions.

\yueying{The theory of constrained random fields was first set forth by \cite{Bertschinger1987}, followed by the first optimisation from \cite{Binney1991}.
\cite{Hoffman1991} then introduced an optimal formalism to construct samples of constrained Gaussian random fields, which was further elaborated and extended by \cite{vandeWeygaert1996}.
In this work, we implement the CR formalism mainly following the work of \cite{vandeWeygaert1996}.}
These authors developed a general class of constraints formatted as convolutions with the linear density field. In this formalism, the CR technique can impose constraints on different characteristics of a Gaussian random field, such as multiple properties of the density peak, the tidal field, as well as the peculiar velocity field. 
These general convolution-type constraints can be applied at arbitrary positions, on different scales, and with various convolution kernels, providing a powerful tool for generating the desired large-scale structures in cosmological simulations.
This method, together with other associated techniques has been used in cosmological hydrodynamic simulations in the past few years to study dark matter halos and galaxy formation \citep[see, e.g.][]{vandeWeygaert1994, Romano2011, Romano2014, Roth2016, Porciani2016, Pontzen2017}.

In this work, we construct a series of constrained ICs in volumes of (20 $\hmpc$)$^3$ with the CR technique. 
We impose a density peak of height $5 \sigma_0 (\rm R_G)$ on scales of $\rm R_G = 1 \hmpc$, with different peak features such as compactness, ellipticity, as well as constraints to the surrounding tidal field strength.
One of the major benefits of the CR technique is that it allows us to investigate the growth of a massive halo (required to be the host of high QSOs) with $M_{\rm halo} \sim 10^{12} M_{\odot}$ at redshift $z \sim 7$, in a relatively small simulation box.
This would otherwise only be achieved (for uniform resolution simulations) with much larger volume normal unconstrained simulations (e.g. 400 $\hmpc$ per side for \textsc{BlueTides}). 
Using this technique, we reduce the computational costs for these rare peaks by a factor of $(400/20)^3 \sim 8000$.

\yn{CR technique provides a general and efficient way to directly study the impact of large scale structure on the formation of the early QSOs and galaxies. It allows us to impose precise controls over multiple large-scale features, such as the height and shape of the density peaks,  as well as the peculiar velocity and tidal field at the site of the peak.
Another common method to simulate a particular region of interest with a given large scale feature is zoom-in simulation, which is designed to completely reproduce a "user-selected" region (for example, a specific halo) from a low-resolution simulation. 
Recent work has also been carried out to combine zoom-in simulation with CR technique to perform high-resolution studies of particular objects with even higher computational efficiency~\citep[e.g.][]{Romano2014,Roth2016}.}

\yueying{Some early works \citep[e.g.][]{Romano2011} has applied a similar setup using the constrained technique in small volume simulations to study some random samples of the $M_{\mathrm{halo}} \sim 10^{12} M_{\odot}$ dark matter halo at $z \sim 6$ as the potential QSO hosts. 
In this work, we carry out a series of constrained simulations where we systematically vary the initial density peak properties. Simulations are run with full baryonic physical models of galaxy formation and SMBH growth and feedback. Therefore, we can directly diagnose the effect of various large-scale features on the early QSO and galaxy growth.}

The paper is organized as follows. 
In Section~\ref{section2:Method}, we review the basic formalism of the CR technique, illustrate how to constrain multiple features of the density peaks in the Gaussian random field. We then introduce the relevant sub-grid physics of the hydrodynamic code \texttt{MP-Gadget} (which we use to carry out the constrained simulations in this study).
In Section~\ref{section3:show-example1} we show two illustrative examples of the application of the CR technique to our small 20 $\hmpc$ simulation box and reconstruct the large scale features to recover the BH growth for the rare SMBHs and their hosts found in \textsc{BlueTides} simulation.
In Section~\ref{section4:result}, we investigate the effects of the different features of the primordial density peak relevant to the early BH growth.
In Section~\ref{section5:Obs}, we look into the relation between SMBHs in our constrained simulations and their host galaxies.
Finally, we summarize and conclude the paper in Section~\ref{section6:conclusion}.

\section{Method}
\label{section2:Method}

\subsection{Formalism of constrained realization}

In this section, we briefly review the CR formalism introduced by \cite{Hoffman1991,vandeWeygaert1996}, which is the foundation of this work. 
We also release our \textsc{gaussianCR} python module  \footnote{\url{https://github.com/yueyingn/gaussianCR}} to implement the CR technique and impose constraints on a random Gaussian realization of IC.

Our goal is to construct a density field realization $f(\mathbf x)$ subject to a set of $M$ constraints:
\begin{equation}
\Gamma = \{ C_i \equiv C_i [f; \mathbf r_c] = c_i; i = 1,...,M \}.
\end{equation}
\yueying{The constraints $C_i[f;\mathbf r_c]$ can be written in a general convolutional format}

\begin{equation}
\label{equation:extract_info}
C_i[f;\mathbf r_c] = \int d\mathbf{x} f(\mathbf x) H_i(\mathbf x, \mathbf r_c) = \int \frac{d \mathbf{k}} {(2 \pi)^3} \hat{f}(\mathbf k) \hat{H}^{*}_i (\mathbf k, \mathbf r_c) = c_i
\end{equation}
\yueying{i.e., for constraint $C_i[f;\mathbf r_c]$, we convolve the $f(\mathbf x)$ field with some kernel $H_i(\mathbf x, \mathbf r_c)$ and impose a specific convolved value $c_i$ at position $\mathbf x = \mathbf r_c$.}
Here $\hat{f}(\mathbf k)$ and $\hat{H}^{*}_i (\mathbf k, \mathbf r_c)$ are the Fourier transforms of $f(\mathbf x)$ and $H_i(\mathbf x, \mathbf r_c)$ respectively, \yueying{with convention of the Fourier transform $f(\mathbf{x}) = \int \frac{d \mathbf{k}}{(2 \pi)^3} \hat{f}(\mathbf k) e^{i\mathbf{k}\cdot\mathbf{x}}$ applied throughout this work.}

Given a certain constraint set $\Gamma$, one can build a corresponding "ensemble mean field" via:
\begin{equation}
\label{equation:Ensemble_mean}
\bar{f_{\Gamma}}(\mathbf x) \equiv \langle f(\mathbf x)|\Gamma \rangle = \xi_i(\mathbf x) \xi^{-1}_{ij} c_j
\end{equation}
where $\xi_i(\mathbf x) = \langle f(\mathbf x)C_i[f;\mathbf r_c] \rangle$ is the cross-correlation between the $f(\mathbf x)$ field and the $i$th constraint $C_i$, and $\xi^{-1}_{ij}$ is the ($ij$)th element of the inverse of constraint's covariance matrix $\langle C_i C_j \rangle$. 
Notice that summation over repeated indices is used in Eq.~\ref{equation:Ensemble_mean}.
The ensemble mean field $\bar{f_{\Gamma}}(\mathbf x)$ can be interpreted as the "most likely" field subject to the set of constraints $\Gamma$.

The CR formalism further introduces the "residual field" $F(\mathbf x) \equiv f(\mathbf x) - \bar{f_{\Gamma}}(\mathbf x)$ as the difference between an arbitrary Gaussian realization $f(\mathbf x)$ satisfying the constraint set $\Gamma$ and the ensemble mean field $\bar{f_{\Gamma}}(\mathbf x)$ of all those fields.
The crucial idea behind the CR construction method is based on the fact that, the complete probability distribution $\mathscr{P} [F|\Gamma]$ of the residual field $F(\mathbf x)$ is independent of the numerical values $c_i$ of the constraints $\Gamma$ \citep[c.f.][for detailed derivations]{Hoffman1991,vandeWeygaert1996}. i.e., for any $\Gamma_1$, $\Gamma_2$, we have
\begin{equation}
\mathscr{P} [F|\Gamma_1] = \mathscr{P} [F|\Gamma_2]
\end{equation}
Therefore, one can construct the desired realization under constraint sets $\Gamma$ by properly sampling a residual field $F(\mathbf x)$ from a random, unconstrained realization $\tilde{f} (\mathbf x)$ and then adding that $F(\mathbf x)$ to the ensemble field $\bar{f_{\Gamma}}(\mathbf x)$ corresponding to $\Gamma$. 
The formalism can be written as:
\begin{equation}
\label{equation:construct_fx}
\begin{split}
f(\mathbf x) & = F(\mathbf x) + \bar{f}_{\Gamma}(\mathbf x) \\
&= (\tilde{f} (\mathbf x) - \bar{f}_{\tilde{\Gamma}}(\mathbf x)) + \bar{f}_{\Gamma}(\mathbf x) \\ 
&= (\tilde{f} (\mathbf x) - \xi_i(\mathbf x) \xi^{-1}_{ij} \tilde{c}_j)  + \xi_i(\mathbf x) \xi^{-1}_{ij} c_j \\
&= \tilde{f} (\mathbf x)  + \xi_i(\mathbf x) \xi^{-1}_{ij} (c_j - \tilde{c}_j)
\end{split}
\end{equation}

i.e., we are treating the original $\tilde{f}(\mathbf x)$ as a field subject to constraint sets $\tilde{\Gamma}$ with value $\tilde{c}_j = C_j [\tilde{f}; \mathbf r_c]$, ($\tilde{c}_j$ is the original value of the unconstrained field), and $\bar{f}_{\tilde{\Gamma}}(\mathbf x) = \xi_i(\mathbf x) \xi^{-1}_{ij} \tilde{c}_j$ is the ensemble mean field corresponding to $\tilde{\Gamma}$.
From $\tilde{f}(\mathbf x) - \bar{f}_{\tilde{\Gamma}}(\mathbf x)$ we get the residual field $F(\mathbf x)$ from a random unconstrained realization, and adding that to $\bar{f}_{\Gamma}(\mathbf x)$ results in the field $f(\mathbf x)$ satisfying constraint $\Gamma$.
It is well established in \cite{vandeWeygaert1996} that the $f(\mathbf x)$ field constructed in this way is a properly sampled realization subject to the desired constraint $\Gamma$.

Using the definition of the matter power spectrum $P(k)$,
\begin{equation}
(2\pi)^3 P(k_1) \delta_D(\mathbf k_1 - \mathbf k_2) = \langle \hat{f}(\mathbf k_1) \hat{f}^{*}(\mathbf k_2) \rangle
\end{equation}
we can write down the formalism for $\xi_i(\mathbf x)$ and $\xi_{ij}$ as follows:

\begin{equation}\label{equation:xi_i}
\begin{split}
\xi_i(\mathbf x) & \equiv \langle f(\mathbf x) C_i[f;\mathbf r_c] \rangle \\ 
& = \langle \int \frac{d \mathbf{k_1}} {(2 \pi)^3} \hat{f}(\mathbf k_1) e^{i \mathbf k_1 \cdot \mathbf x}  \int \frac{d \mathbf{k_2}} {(2 \pi)^3} \hat{f}^{*}(\mathbf k_2) \hat{H}_i (\mathbf k_2,\mathbf r_c) \rangle \\
&= \int \frac{d \mathbf{k_1}} {(2 \pi)^3} \frac{d \mathbf{k_2}} {(2 \pi)^3} \langle \hat{f}(\mathbf k_1) \hat{f}^{*}(\mathbf k_2)\rangle \hat{H}_i (\mathbf k_2, \mathbf r_c) e^{i \mathbf k_1 \cdot \mathbf x} \\
&= \int \frac{d \mathbf{k}} {(2 \pi)^3} P(k) \hat{H}_i(\mathbf k, \mathbf r_c) e^{i \mathbf{k} \cdot \mathbf x}
\end{split}
\end{equation}

and 

\begin{equation}\label{equation:xi_ij}
\begin{split}
\xi_{ij} & \equiv \langle C_i[f;\mathbf r_c] C_j[f;\mathbf r_c] \rangle \\
&= \langle \int \frac{d \mathbf{k_1}} {(2 \pi)^3} \hat{f}(\mathbf k_1) \hat{H}^{*}_i (\mathbf k_1, \mathbf r_c) \int \frac{d \mathbf{k_2}} {(2 \pi)^3} \hat{f^*}(\mathbf k_2) \hat{H}_j (\mathbf k_2, \mathbf r_c) \rangle \\
&= \int \frac{d \mathbf{k}} {(2 \pi)^3} \hat{H}^{*}_i (\mathbf k, \mathbf r_c) \hat{H}_j (\mathbf k, \mathbf r_c) P(k)
\end{split}
\end{equation}

Plugging Eq.~\ref{equation:xi_i} and Eq.~\ref{equation:xi_ij} into Eq.~\ref{equation:Ensemble_mean},
the desired constrained field can be constructed via 
\begin{equation}
\label{equation:final_expression}
f(\mathbf x) = \int \frac{d \mathbf{k}} {(2 \pi)^3} [\hat{\tilde{f}}(\mathbf k) + P(k) \hat{H}_i(\mathbf k, \mathbf r_c) \xi^{-1}_{ij} (c_j - \tilde{c}_j)] e^{i \mathbf{k} \cdot \mathbf x}
\end{equation}


\begin{figure*}
\begin{center}
\includegraphics[width=1.8\columnwidth]{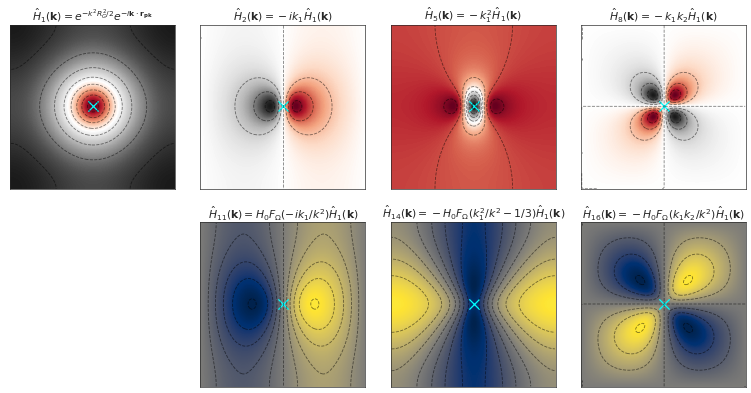}
\end{center}
    \caption{Illustration of a subset of the 18 constraints imposed on the density peak. 
    The panels show the $\xi_i (\bf x)$ field constructed via Eq.~\ref{equation:xi_i} projected into the xy plane. The $\hat{H}(\bf k)$ kernels have a Gaussian smoothing scale of $\rm R_G = 1$ $\hmpc$. The peak position $\mathbf{r}_\mathrm{pk}$ is located at the centre of the box. The box is 20 $\hmpc$ per side.
    \textit{Top panels:} Illustration of the constraint kernels corresponding to the zeroth ($\hat{H}_1$), first ($\hat{H}_2$) and second order ($\hat{H}_{5,8}$) derivatives of the density peak, which shapes the immediate surrounding of the density peak.
    \textit{Bottom panels:} Illustration of the constraint kernels corresponding to the first ($\hat{H}_{11}$) and second ($\hat{H}_{14,16}$) order derivatives of the gravitational field (i.e., the peculiar velocity and the tidal field). These constraints sculpt the large scale features of the peak's surroundings.}
    \label{fig:H18_subset}
\end{figure*}

\begin{figure*}
\begin{center}
\includegraphics[width=2.1\columnwidth]{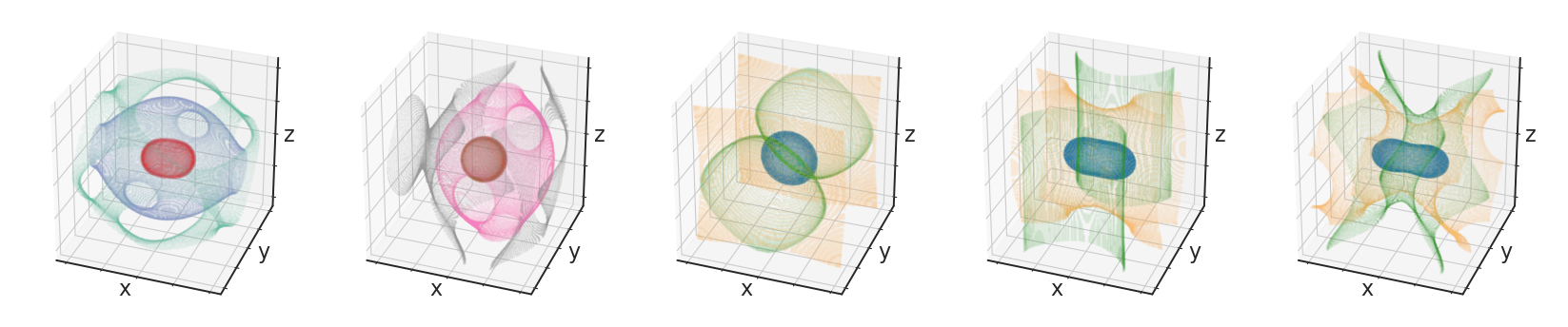}
\end{center}
    \caption{3D contour plot of the density field illustrating the effect of imposing constraints on the second order derivatives of the density peak $\partial^2{f_\mathrm{G}}/{\partial x_i \partial x_j}$ (the first panel), the peculiar velocity field $V_{\mathrm G}$ (the second panel), and the tidal field $T_{ij}$ (the right three panels). 
    \yueying{The three layers in each panel represent the isodensity contours with density values equal to 98\%, 60\% and 30\% of the overall density distribution of the box.}
    \textbf{\textit{First panel:}} Illustration of the mass ellipsoid of the density peak. We show here the ensemble mean field of a $3 \sigma_0$ peak constructed via $\hat{H}_1 (\bf k)$ and second order derivatives of $\hat{H}_5 (\bf k)$ $\sim \hat{H}_{10} (\bf k)$, giving a triaxial peak positioned at the centre of the box, with axial ratio 4:3:2 in the x,y and z directions respectively. The principal axes of the mass ellipsoid align with the coordinate axes.
    \textbf{\textit{Second panel:}} Ensemble mean field constructed with $\hat{H}_1 (\bf k)$ and $\hat{H}_{11} (\bf k)$, showing a $3 \sigma_0$ density peak with peculiar velocity $v_x = 60 \kms$ ($+x$ direction) on scale of 1 $\hmpc$.
    \textbf{\textit{Right three panels:}} Illustration of the ensemble mean field constructed via $\hat{H}_1 (\bf k)$ and $\hat{H}_{14} (\bf k)$ $\sim \hat{H}_{18} (\bf k)$,
    showing a $3 \sigma$ density peak subject to a tidal field with shear magnitude $\epsilon$ = 60 $\kmsmpc$ and three differing shear angles, $\omega = 1 \pi$ (the 3rd panel), $\omega = 1.5 \pi$ (the 4th panel) and $\omega = 2 \pi$ (the 5th panel).
    The principal axes (eigenvectors) of the tidal field are aligned with the coordinates axes. See text for further explanation.}
    \label{fig:contour3D}
\end{figure*}


\subsection{Constraint parameters of a density field}
\label{subsection:2.2}
We use the prescription introduced in \cite{vandeWeygaert1996} to construct a peak in the smoothed density field at an arbitrary position $\mathbf{r}_\mathrm{pk}$.
Apart from the location, scale and the height of the peak, one can also impose constraints on the first and second-order derivative of the density field $f(\mathbf x)$ to shape the immediate surrounding of the peak, as well as impose constraints on the tidal field and peculiar velocity field at $\mathbf{r}_\mathrm{pk}$ to sculpt the global matter distribution around the peak.
\yueying{Throughout this work, we apply the constraints covariantly on the same position $\mathbf{r}_\mathrm{pk}$, therefore, for simplicity we omit $\mathbf{r}_\mathrm{pk}$ and use the notation $ \hat{H}(\mathbf{k}) = \hat{H}(\bf k, \mathbf{r}_\mathrm{pk})$ in the following sections.}

Here we briefly introduce the 18 constraints used to characterize the peak in the smoothed density field (i.e., $i$=1,2,..18 in Eq.~\ref{equation:final_expression}).
From Eq.~\ref{equation:Ensemble_mean}, we can see that the ensemble mean field is effectively the superposition of the constraint correlation fields $\xi_i(\mathbf x)$ weighted by a factor $\Sigma_j \xi^{-1}_{ij} c_j$.
To illustrate how these constraints work, we plot in Appendix~\ref{Appendix C} the $\xi_i (\mathbf x)$ fields constructed via Eq.~\ref{equation:xi_i} with the 18 $k$-space convolution kernels $\hat{H}_1 (\bf k)$ $\sim \hat{H}_{18} (\bf k)$ imposing on position $\mathbf{r}_\mathrm{pk}$ respectively.
In Figure~\ref{fig:H18_subset}, we show a subset of those $\xi_i (\mathbf x)$ fields constructed using their $\hat{H}_i (\bf k)$, as a representation of the different types of kernels. 
The panels show the projection of the $\xi_i(\bf x)$ field onto the $xy$ plane, with the box size 20 $\hmpc$ per side and peak position $\mathbf{r}_\mathrm{pk}$ located at the centre of the box.
The Gaussian kernel $\hat{H}(\bf k)$ has a characteristic scale of $R_\mathrm{G} = 1$ $\hmpc$.
Note that the periodic boundary condition is automatically imposed by the Fourier transform.
The expressions for each of the $\hat{H} (\bf k)$ kernels are given on top of each panel.

The 18 constraints $C_1 \sim C_{18}$ can be divided into five categories, which we briefly summarize as follows.

\begin{enumerate}
	\item $f_{0,\rm G}$, (corresponding to $C_1$,) is the zeroth order Gaussian smoothed density field $f_\mathrm{G}(\bf x)$.
	with 
	\begin{equation}
	   \hat{H}_1 (\mathbf{k}) = e^{-k^2R^{2}_\mathrm{G}/2}e^{-i \mathbf{k} \cdot \mathbf{r_{pk}}}
	\end{equation}
	$C_1$ specifies the characteristic scale $\rm R_\mathrm{G}$ of the Gaussian kernel, the location $\mathbf{r}_\mathrm{pk}$ of the peak, and the height of the peak  $c_1 = \nu_c \sigma_0$ in the smoothed density field $f_\mathrm{G}(\bf x)$.
	
	The unit of peak height $\sigma_0$ is the variance of $f_G(\bf x)$, with
	\begin{equation}
	\label{equation:sigma0}
    \sigma^2_{0}(R_{\mathrm{G}}) = \langle f_\mathrm{G}(\mathbf{x}) f_\mathrm{G}(\mathbf{x}) \rangle
    = \int \frac{d\mathbf{k}}{(2\pi)^3} P(k) \hat{W}^{2}(kR_{\mathrm{G}}) 
    \end{equation}
    \yueying{where $\hat{W} (kR_{\mathrm{G}}) = \exp(-k^2R_{\mathrm{G}}^2/2)$ is the Fourier transform of the Gaussian kernel.}
	
	Note that $\sigma_0$ is the zeroth order spectral moment, where the $l$th spectral moment is defined as
	\begin{equation}
	\label{equation:spectral_moment}
    \sigma^2_{l}(R_{\mathrm{G}}) \equiv \int \frac{d\mathbf{k}}{(2\pi)^3} P(k) \hat{W}^{2}(kR_{\mathrm{G}}) k^{2l} 
    \end{equation}
	and can be used to evaluate the variance of derivatives of the $f_\mathrm{G}(\bf x)$ field.

	\item $f_{1,\rm G}$, (corresponding to $C_2 \sim C_4$,) are the first order derivatives of the $f_{\rm G}(\bf x)$ field.
	with 
	\begin{equation}
	   \hat{H}_j (\mathbf {k}) = -i k_i \hat{H}_1(\mathbf{k})
	\end{equation}
	(where $j = 2,3,4$ and $i = j-1$), $C_2 \sim C_4$ are imposed on the $\partial{f_G}/\partial x_i$ at $\mathbf{r}_\mathrm{pk}$ (in the x,y and z directions respectively). 
	Throughout this study, we set $c_2 \sim c_4$ to be zero, and therefore ensure that $\mathbf{r}_\mathrm{pk}$ is the maximum of the peak.
	
	\item $f_{2,\rm G}$, (corresponding to $C_5 \sim C_{10}$,) are the second order derivatives of the smoothed density field,
	with 
	\begin{equation}
	   \hat{H}_l (\mathbf k) = -k_i k_j \hat{H}_1(\mathbf{k})
	\end{equation}
	(where $l = 5,...,10$ and $(i,j)$ = (1,1),(2,2),(3,3),(1,2),(1,3),(2,3), $C_5 \sim C_{10}$ constrain the $\partial^2{f_G}/{\partial x_i \partial x_j}$ with the six constraint values $c_5 \sim c_{10}$ corresponding to the diagonal and off-diagonal components of the matrix.
	Since $\partial^2{f_G}/{\partial x_i \partial x_j}$ should be negative definite at $\mathbf{r}_\mathrm{pk}$, and the vicinity of the density peak is ellipsoidal, 
	we can transform $c_5 \sim c_{10}$ to a set of more physical quantities: the compactness $\rm x_d$, the two axial ratios $a_{12} \equiv a_1/a_2$ and $a_{13} \equiv a_1/a_3$, and the orientation of the peak specified by the three Euler angles $\alpha_1$, $\beta_1$, $\gamma_1$ via:
	\begin{equation}
	\frac{\partial^2{f_G}}{\partial x_i \partial x_j} = -\sum_{k=1}^{3} \lambda_k A_{ki} A_{kj}, \quad i,j = 1,2,3
	\label{equation:ellip}
	\end{equation}
	with 
	\begin{equation}
	\lambda_1 = \frac{x_d \sigma_2(\mathrm{R_G})}{1+a^2_{12}+a^2_{13}}, \quad \lambda_2 = \lambda_1 a^2_{12}, \quad \lambda_3 = \lambda_1 a^2_{13}.
	\end{equation}
	
	Here $A_{ij}$ is the transformation matrix determined by the three Euler angles $\alpha_1$, $\beta_1$, $\gamma_1$ to rotate from the original coordinate to the principal axes of the mass ellipsoid.
	Throughout this work, the compactness of the density peak $\rm x_d$ is given in unit of $\sigma_2(\mathrm{R_G}) = \langle \nabla^2 f_G\nabla^2 f_G \rangle ^{1/2}$ which is the second order spectral moment, 
	
	As an illustration of the ellipticity, we plot in the first panel of Figure~\ref{fig:contour3D} the 3D contour plot of the ensemble mean field constructed via $\hat{H}_1 (\bf k)$ and $\hat{H}_5 (\bf k)$ $\sim \hat{H}_{10} (\bf k)$ using Eq.~\ref{equation:Ensemble_mean}. 
	The coloured contours show the three isodensity surfaces of the triaxial density peak. 
	The peak is positioned at the centre of the box, with principal axes aligned with the coordinate axes and axial ratio 4:3:2 for the x,y and z directions respectively.
	
	\item $V_G$, (corresponding to $C_{11} \sim C_{13}$): constrain the (smoothed) peculiar velocity of the field. 
	In the linear regime, the peculiar velocity is induced by the gravitational acceleration that corresponds to the first order derivative of the gravitational potential.
	The peculiar velocity field is related to the matter density field via
	\begin{equation}
	    \mathbf{v_k} = H_0 F({\Omega_0}) \frac{i\mathbf{k}}{k^2} \delta_{\mathbf k}
	\end{equation}
	where $H_0$ is the Hubble constant, and $F({\Omega_0}) = {\Omega_0}^{0.6}$.
	We can impose constraints on the smoothed peculiar velocity field at the location of the peak $c_{11}=v_x$, $c_{12}=v_y$, $c_{13}=v_z$ in unit of $\kms$
	with 
	\begin{equation}
	\hat{H}_{j} (\mathbf{k}) = H_0 F({\Omega_0}) (-ik_i/k^2) \hat{H}_1(\mathbf{k})
	\end{equation}
	(where $j=11,12,13$, $i=1,2,3$). 
	
	Note that the factor of $k^{-2}$ in the $\hat{H}_{j}$ kernel indicates that the peculiar velocity field is generated by the large scale matter distribution.
	In the second panel of Figure~\ref{fig:contour3D}, we plot an ensemble mean field constructed with $\hat{H}_1 (\bf k)$ and $\hat{H}_{11} (\bf k)$, showing a $3 \sigma_0$ density peak with peculiar velocity $v_x = 60 \kms$ on scale of 1 $\hmpc$. 
	The density contours illustrate that the peculiar velocity is induced by the asymmetry of the matter density distribution in the $x$ direction.  The overdensity to the right of the peak ($+x$ direction) attracts the matter from the left, and therefore gravitationally accelerates the density peak in the positive $x$ direction.
	
	\item $T_G$, (corresponding to $C_{14} \sim C_{18}$): constrain the tidal field at the site of the density peak induced by the second order derivatives of the gravitational field. In the linear regime, it is analogous to the shear field.
	with 
	\begin{equation}
	\hat{H}_{l} (\mathbf{k}) = - H_0 F({\Omega_0}) (k_i k_j/k^2 - 1/3 \delta_{ij}) \hat{H}_1(\mathbf{k})
	\end{equation}
	(where $l = 14,...18$ and $(i,j)=(1,1),(2,2),(1,2),(1,3),(2,3)$), $c_{14} \sim c_{18}$ corresponds to the traceless tidal tensor component $T_{G,ij}$ of the density peak, in unit of $\kmsmpc$. 
	Following \cite{vandeWeygaert1996}, we can parametrize the tidal field via {$\epsilon$, $\omega$, $\alpha_2$, $\beta_2$, $\gamma_2$}, where one parameter $\epsilon$ sets the total magnitude of the tidal field,  $\omega$ distributes the relative strength of the tidal field along the three principal axes, and three Euler angles set the orientation of the principal axes:
	\begin{equation}
	T_{G,ij} = \sum_{k=1}^{3} \zeta_k Q_{ki} Q_{kj}
	\end{equation}
	$Q_{ij}$ is the transformation matrix determined by $\alpha_2$, $\beta_2$, $\gamma_2$, and
	\begin{equation}
	\label{equation:shear_angle}
	\zeta = [\epsilon\cos(\frac{\omega + 2\pi}{3}), \epsilon\cos(\frac{\omega - 2\pi}{3}), \epsilon\cos(\frac{\omega}{3})]
	\end{equation}
	are the three eigenvalues of the tidal tensor, with positive values indicating dilation and negative values indicating contraction. 
	
	As a further illustration of the tidal field, we plot in the right three panels of Figure~\ref{fig:contour3D} the 3D contour plot of the ensemble mean field constructed via $\hat{H}_1 (\bf k)$ and $\hat{H}_{14} (\bf k)$ $\sim \hat{H}_{18} (\bf k)$ with Eq.~\ref{equation:Ensemble_mean}, showing a $3 \sigma_0$ density peak subject to a tidal field with magnitude $\epsilon = 60 \kmsmpc$ and with different angles $\omega$: $\omega = 1 \pi$ (the 3rd panel), $\omega = 1.5 \pi$ (the 4th panel) and $\omega = 2 \pi$ (the 5th panel).
    The principal axes (eigenvectors) of the tidal field are aligned with the coordinates. 
    For $\omega = 1 \pi$, the three eigenvalues of the tidal tensor are [30 -60 30] $\kmsmpc$ (c.f. Eq.~\ref{equation:shear_angle}) in the x, y and z directions \yy{respectively}, and therefore the density peak is equally elongated along both x and z axes and is compressed in the y direction.
    For $\omega = 1.5 \pi$, the tidal tensor has eigenvalues of [52 -52 0] $\kmsmpc$ in the x, y and z directions, and the density peak is elongated along the x axis, compressed along the y axis, and not affected in the $z$ axis direction. 
    For $\omega = 2 \pi$, the tidal tensor has eigenvalues of [60 -30 -30] $\kmsmpc$ in the x, y and z directions, and therefore the density peak is elongated along the x axis and is equally compressed in the y and z directions.
    
\end{enumerate}

To briefly summarise, the first 10 constraints ($C_1 \sim C_{10}$) together determine the density distribution in the immediate vicinity of the density peak. 
$C_{11} \sim C_{18}$ are imposed on the derivatives of the local gravitational potential (the peculiar velocity and the tidal field) at the peak position. They sculpt the global matter distribution on larger scales since the gravitational potential perturbation is the weighted sum of all density perturbations throughout the universe.

With proper parametrization introduced above, one can use 15 physical parameter sets $\{$ $\nu$,$x_d$,$a_{12}$,$a_{13}$,$\alpha_1$,$\beta_1$,$\gamma_1$, $v_x$,$v_y$,$v_z$,$\epsilon$, $\omega$,$\alpha_2$,$\beta_2$,$\gamma_2$ $\}$ to characterize a density peak in the smoothed density field. (Note that we always set the three first derivatives of the peak $c_2 \sim c_4$ to be zero).
In Appendix~\ref{Appendix B}, we give some further illustration by showing the ensemble mean field of density peaks with varying peak parameters.

\subsection{Simulation Set up}

We use the massively parallel cosmological smoothed particle hydrodynamic (SPH) simulation software, \texttt{MP-Gadget} \citep{Feng2016}, to run all the simulations in this paper. 
The hydrodynamics solver of \texttt{MP-Gadget} adopts the new pressure-entropy formulation of SPH \citep{Hopkins2013}.
We also apply a variety of sub-grid models to model the galaxy and black hole formation and associated feedback processes.
In the simulations, gas is allowed to cool through both radiative processes~\citep{Katz} and metal cooling. 
The metal cooling rate is obtained by scaling a solar metallicity template according to the metallicity of gas particles, following the method described in \cite{Vogelsberger2014}.
Star formation (SF) is based on a multi-phase SF model ~\citep{SH03} with modifications following~\cite{Vogelsberger2013}.
\yn{We use a sub-grid $H_2$-based star formation rate following the prescription of \cite{Krumholtz}, that estimate the fraction of molecular hydrogen gas from the baryon column density, and in turn couples the density gradient to the SF rate.
This effectively models the SF in the (unresolved) molecular phase of ISM in cosmological simulations.}
Type II supernova wind feedback (the model used in Illustris ~\citep{Nelson}) is included, assuming wind speeds proportional to the local one dimensional dark matter velocity dispersion. 

In our simulation, we model BH growth and AGN feedback in the same way as in the \textit{MassiveBlack} $I \& II$ simulations, using the BH sub-grid model developed in \cite{SDH2005,DSH2005} with modifications consistent with \textsc{BlueTides}. 
BHs are seeded with an initial seed mass of $M_{\mathrm {seed}} = 5 \times 10^5 h^{-1} M_{\odot}$ (commensurate with the resolution of the simulation) in halos with mass more than $5 \times 10^{10} h^{-1} M_{\odot}$. 
The gas accretion rate onto BH is given by Bondi accretion rate,
\begin{equation}
\label{equation:Bondi}
    \dot{M}_{\rm B} = \frac{4 \pi G^2 M_{\rm BH}^2 \rho}{(c^2_s+v_{\rm rel}^2)^{3/2}}
\end{equation}
where $c_s$ and $\rho$ are the local sound speed and density of the cold gas, $v_{\rm rel}$ is the relative velocity of the BH to the nearby gas. 

We allow for super-Eddington accretion in the simulation but limit the accretion rate to 2 times the Eddington accretion rate:
\begin{equation}
\label{equation:Meddington}
    \dot{M}_{\rm Edd} = \frac{4 \pi G M_{\rm BH} m_p}{\eta \sigma_{T} c}
\end{equation}
where $m_p$ is the proton mass, $\sigma_T$ the Thompson cross section, $c$ is the speed of light, and $\eta=0.1$ is the radiative efficiency of the accretion flow onto the BH.
Therefore, the BH accretion rate is determined by \yy{(same as \textsc{BlueTides} simulation)}:
\begin{equation}
    \dot{M}_{\rm BH} = {\rm Min} (\dot{M}_{\rm B}, 2\dot{M}_{\rm Edd})
\end{equation}
The Eddington ratio defined as $\lambda_{\mathrm {Edd}}= \dot{M}_{\mathrm {BH}}$/2$\dot{M}_{\mathrm {Edd}}$ would usually range from $0 \sim 1$ during the evolution of AGN.

The SMBH is assumed to radiate with a bolometric luminosity $L_{\rm Bol}$ proportional to the accretion rate $\dot{M}_{\rm BH}$:
\begin{equation}
    L_{\rm Bol} = \eta \dot{M}_{\rm BH} c^2
\end{equation}
with $\eta = 0.1$ being the mass-to-light conversion efficiency in an accretion disk according to \cite{Shakura1973}.
5\% of the radiation energy is thermally coupled to the surrounding gas that resides within twice the radius of the SPH smoothing kernel of the BH particle. This scale is typically about 1\% $\sim$ 3\% of the virial radius of the halo.
The AGN feedback energy only appears in kinetic form through the action of this thermal energy deposition, and no other coupling (e.g. radiation pressure) is included. 

All the simulations in this work are run with $2 \times 352^3$ particles in a box of size 20 $\hmpc$. 
The effect of the box size will be further discussed in Appendix~\ref{Appendix A}. 
The cosmological parameters used are from the nine-year Wilkinson Microwave Anisotropy Probe (WMAP) \citep{Hinshaw2013} ($\Omega_0=0.2814$, $\Omega_\Lambda=0.7186$, $\Omega_{\rm b}=0.0464$, $\sigma_8=0.82$, $h=0.697$, $n_s=0.971$).
The resolution of the simulation is same as \textsc{BlueTides}, with $M_{\rm DM} = 1.2 \times 10^7 \rm M_\odot/h$ and $M_{\rm gas} = 2.4 \times 10^6 \rm M_\odot/h$ in the initial conditions.
The mass of a star particle is $M_{*} = 1/4 M_{\rm gas} = 6 \times 10^5 \rm M_\odot/h$. 
The gravitational softening length is 1.8 $\hckpc$ for both DM and gas particles.

\begin{figure*}
\begin{center}
\includegraphics[width=1.9\columnwidth]{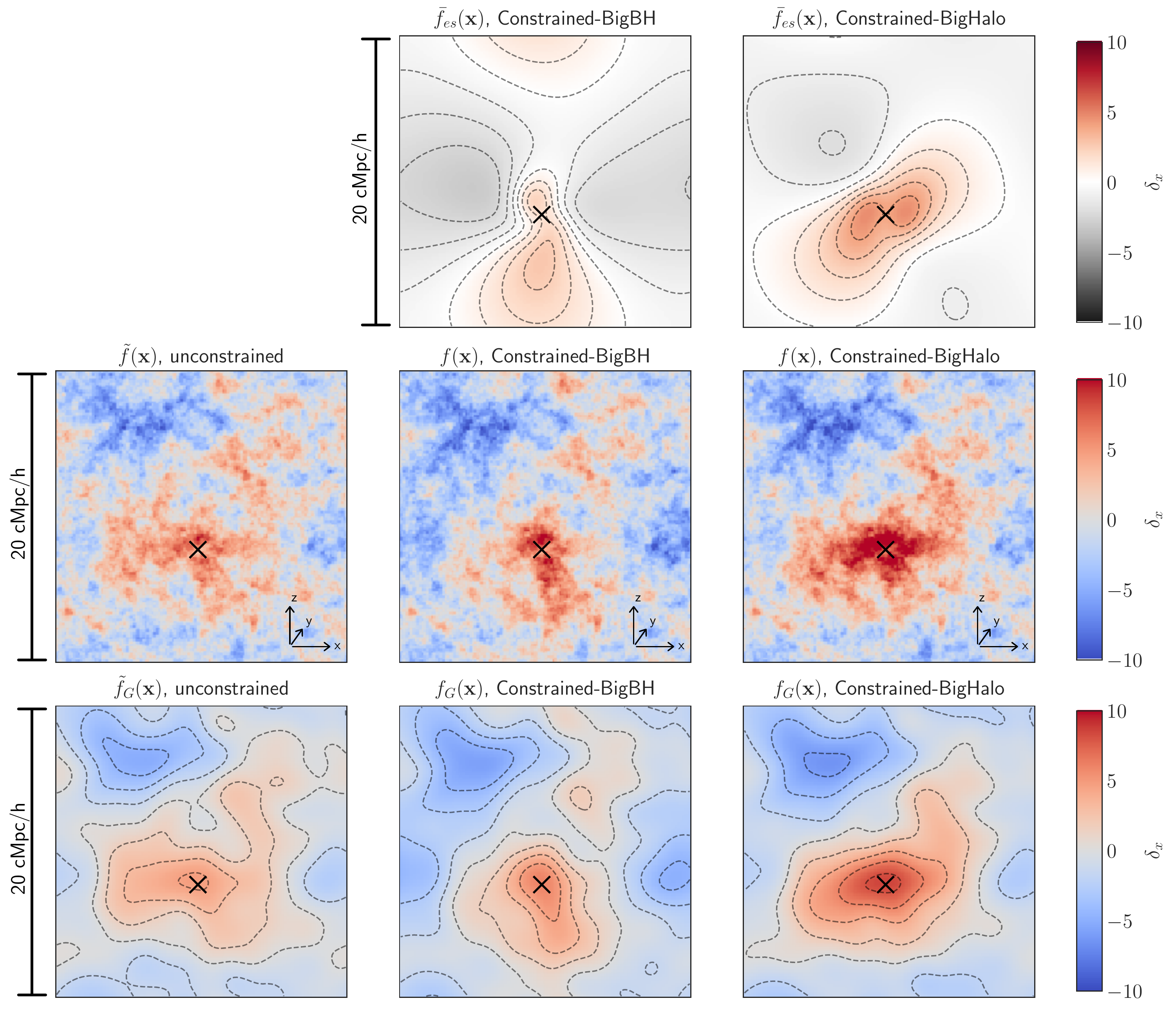}
\end{center}
    \caption{
    Illustration of how to construct the constrained IC given a certain set of peak parameters. 
    Each panel in this plot is 20 $\hmpc$ per side, giving the density contrast field projected onto the $xz$ plane with a slab thickness of 5 $\hmpc$.
    The black crosses mark the position where the constraints are imposed.
    The first panel in the leftmost column shows the density \yy{contrast} field of a random unconstrained realization $\tilde{f} (\mathbf{x})$, with the left bottom panel giving the smoothed $\tilde{f} (\mathbf{x})$ field obtained by convolving with a Gaussian kernel of radius $\rm R_G = 1$ $\hmpc$. 
    The next two columns show the results of imposing the constraints with parameter sets of \textsc{BigBH} (middle) and \textsc{BigHalo} (right) to the unconstrained $\tilde{f} (\mathbf{x})$ separately.
    The middle row shows the two density fields after the constraints, constructed via $f(\mathbf{x})$ = $\tilde{f}(\mathbf{x})$ + $\bar{f}_{\rm es}(\bf x)$, with the ensemble mean field $\bar{f}_{\rm es}(\mathbf{x}) = \xi_i(\mathbf{x}) \xi^{-1}_{ij} (c_j - \tilde{c}_j)$ shown in the top two panels.
    The bottom panels again show the constrained field smoothed with Gaussian kernel to better reveal the large scale density distribution.}
    \label{fig:Constrained-IC}
\end{figure*}

\begin{figure*}
\begin{center}
\includegraphics[width=2\columnwidth]{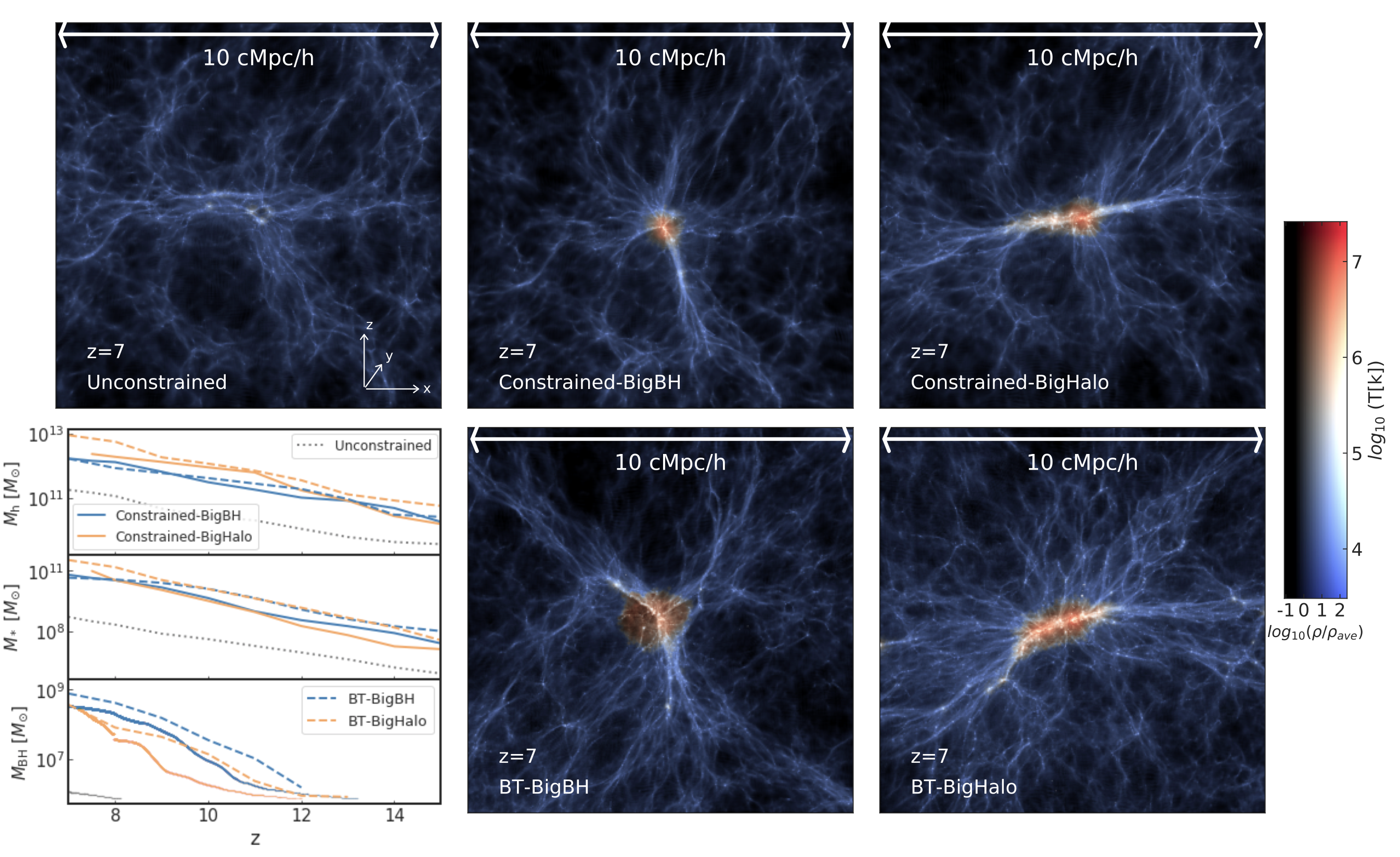}
\end{center}
    \caption{Illustration of the gas density field around the SMBH at $z=7$. 
    \yn{For high density regions, the colour hue is set by the averaged gas temperature, from red to blue indicating warm to cold, as shown by the colour bar aside.}
    Each panel is 10 $\hmpc$ per side and with a slab thickness of 3 $\hmpc$.
    The top left panel is the result of simulation from unconstrained IC.
    The top middle and top right panels are the simulations with constrained IC of \textsc{BigBH} and \textsc{BigHalo}, using the peak parameter sets extracted from IC of \textsc{BlueTides} simulation. 
    For comparison, the two bottom right panels show the corresponding gas density fields around \textsc{BigBH} and \textsc{BigHalo} in \textsc{BlueTides} simulation at $z=7$.
    The left bottom panel gives the growth history of the host halo mass (the upper panel), host galaxy mass (middle panel), and the BH mass (lower panel) as a function of redshift.
    \yueying{The grey line in the three panels give the corresponding growth history from the unconstrained simulation}, the blue and orange solid lines correspond to \textsc{BigBH} and \textsc{BigHalo} from our constrained simulations. As for comparison, the blue and orange dashed lines give the growth history of \textsc{BigBH} and \textsc{BigHalo} in \textsc{BlueTides} simulation.}
\label{fig:density-field-BH0BH3}
\end{figure*}

\begin{figure*}
\begin{center}
\includegraphics[width=1.9\columnwidth]{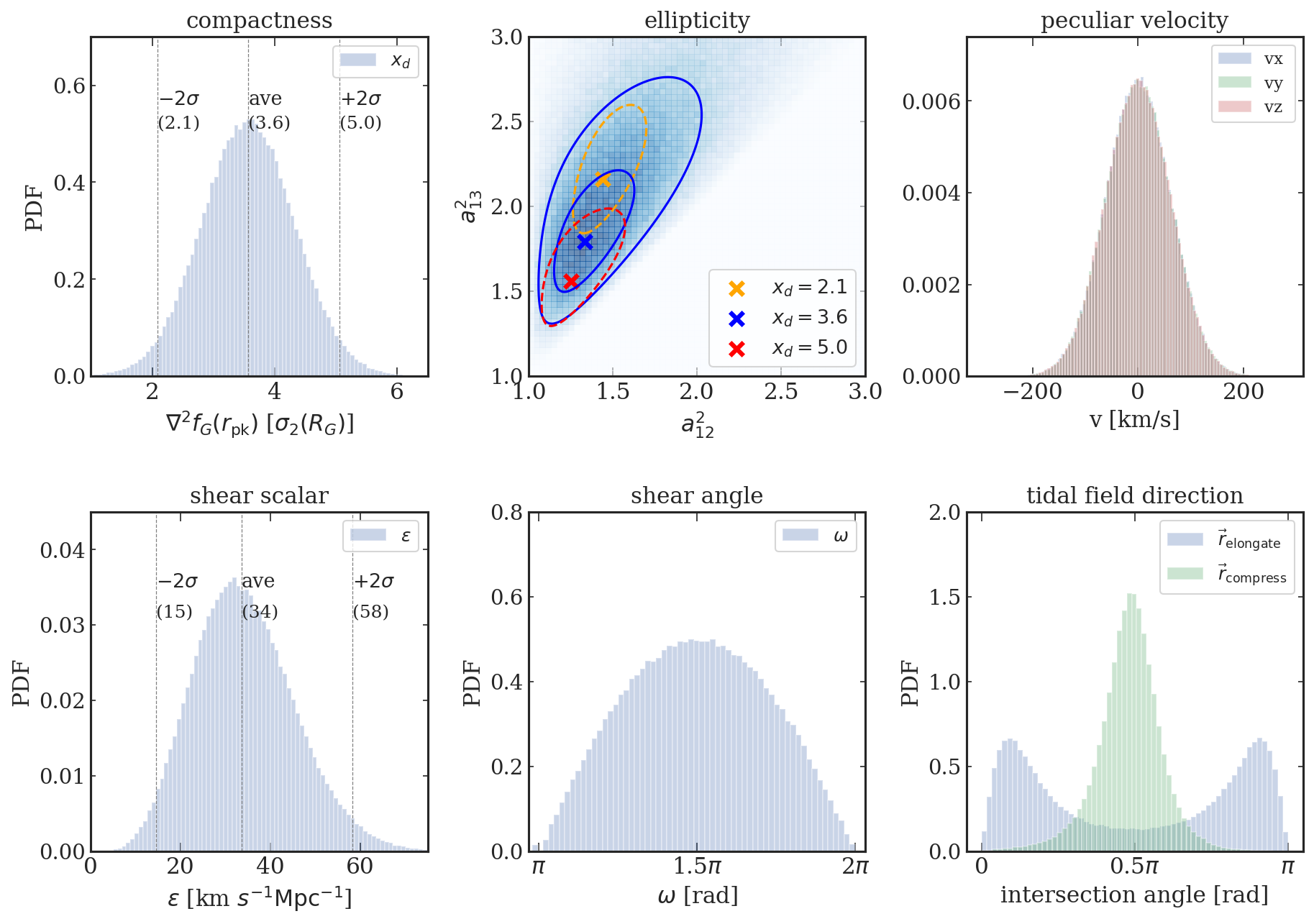}
\end{center}
    \caption{Distribution of the peak parameters conditioned on peak height of $\nu = 5 \sigma_0$. The distribution is extracted from imposing the $5 \sigma_0$ peak constraint on $10^5$ random realizations of initial conditions with simulation boxsize of 20 $\hmpc$ (see text for more details). 
    \textit{\bf{Top left panel}}: Distribution of compactness $\rm x_d$ in units of $\sigma_2 (\rm R_G)$. 
    \textit{\bf{Top middle panel}}: 2D probability distribution of the ellipticity of the density peak $P(a_{12}^2, a_{13}^2 | \rm x_d)$. The blue data points show the $a^2_{12}$, $a^2_{13}$ distribution of the constrained density peaks with $\rm x_d = 3.6$. The blue solid lines show the contours of 97.5\% and 90\% for $P(a_{12}^2, a_{13}^2 | \rm x_d = 3.6)$.
    The orange, blue and red crosses mark the largest probability for $\rm x_d = 2.1$, 3.6 and 5.0, with the orange and red lines giving the 97.5\% contours correspondingly.
    \textit{\bf{Top right panel}}: Distribution of the peculiar velocity of the peak in three directions $v_x$ (blue colour), $v_y$ (green colour) and $v_z$ (red colour) in unit of $\kms$.  
    \textit{\bf{Bottom left panel}}: Distribution of the tidal field magnitude (shear scalar) $\epsilon$ in unit of $\kmsmpc$.
    \textit{\bf{Bottom middle panel}}: Distribution of the shear angle $\omega$ that distributes the relative strength of the tidal field along the three principal axes.
    The $\omega$ distribution peaks at $\omega = 1.5 \pi$, corresponding to the case for which the three eigenvalues of the tidal tensor have signs of (-,0,+). 
    \textit{\bf{Bottom right panel}}: Illustration of how the direction of the tidal field tends to align with the mass ellipsoid of the peak. The green colour represents the distribution of the intersection angle between the long axis of the mass ellipsoid and the direction of the compression of the tidal field, while the blue colour gives the distribution of the intersection angle between the long axis of the mass ellipsoid and the elongating direction of the tidal field.}
\label{fig:param-distribution}
\end{figure*}


\section{Constrained Simulations that resemble \textsc{BlueTides} high-$\sigma$ peaks and their BHs}
\label{section3:show-example1}

\subsection{Building a constrained IC}

In this section, we describe the use of the CR technique to construct density peaks (characterized by the full 18 constraint parameters as described in Section~\ref{subsection:2.2}) in the context of an arbitrary random realization. 
As a proof of concept, here we start by illustrating the procedure by reconstructing the large scale features of two characteristic regions selected from the large volume \textsc{BlueTides} simulation. 
The \textsc{BlueTides} simulation is a cosmological hydrodynamic simulation with a volume of 400 $\hmpc$ and high resolution ($2 \times 7040^3$ particles), run to study the formation and growth of the first massive galaxies and quasars at high redshift, $z>6$ \citep{Feng2016}.
We will show how the density field in a small (constrained) simulation box can be set up to reconstruct some rare high density regions that would otherwise only be realized in a simulation with a much larger volume. 

We select two particular regions in the density field of the large BT volume. (i) \textsc{BT-BigBH} contains the earliest massive BH (i.e. highest redshift quasar),
and (ii) \textsc{BT-BigHalo} the region with the most massive halo.

In \textsc{BT-BigBH}, the central BH reaches a mass  of $4 \times 10^8 M_{\odot}$ at $z=8$: this is the largest BH in the entire volume and it resides in a halo with mass  $8.5 \times 10^{11} M_{\odot}$.
On the other hand, \textsc{BT-BigHalo} is the region with the most massive halo, of mass $5.5 \times 10^{12} M_{\odot}$ and hosts a less massive BH with $M_{\rm BH} = 8 \times 10^7 M_{\odot}$ at $z=8$ (which however does grow to a similar mass as the \textsc{BT-BigBH} by $z=7$).

To select the scale on which we impose the constraints (see Section~\ref{subsection:2.2}) of the density peak, we choose $\rm R_G = 1$ $\hmpc$, which roughly corresponds to the mass of $10^{12} M_{\odot}/h$ within a Gaussian filter:
\begin{equation}
 M(R) \equiv \bar{\rho}V(R) = \bar{\rho}(2\pi)^{3/2}R^3,
\end{equation}
where $\bar{\rho}$ is the mean matter density of the universe.
Theoretical prediction estimates that formation of a halo with mass $10^{12} M_{\odot}/h$ at $z = 8$ corresponds to a peak height of $\nu = 5 \sigma_0$ in the primordial linear density field \citep[see, e.g.][]{Barkana2001}. 
Throughout this study, we will be referring to $5 \sigma_0$ peaks on a scale of $\rm R_G = 1$ $\hmpc$ for most of our constrained simulations.

The procedure to construct a constrained IC goes as follows:
\begin{enumerate}

\item Obtain the parameter set for the selected density peak regions from the large volume \textsc{BlueTides} simulation, what we call \textsc {BT-BigBH} and \textsc{BT-BigHalo}:
    \begin{enumerate}
        \item Extract the progenitor region in the IC (of size 20 $\hmpc$ on a side) of these two targets.
                
        \item Linearly extrapolates the primordial density field to $z=0$ and find the peak position in the Gaussian smoothed field. 
        
        \item Convolve the density field with the 18 kernels $\hat{H}_1 (\bf k)$ $\sim \hat{H}_{18} (\bf k)$ (c.f. Eq.~\ref{equation:extract_info}) with $\rm R_G = 1$ $\hmpc$, to extract the peak parameters for the constrained parameter sets \{$c_j$\}.
    \end{enumerate}

\item Impose the constraint.
    \begin{enumerate}
        \item Generate a random realization $\tilde{f} (\mathbf{x})$. Choose an arbitrary position to impose the peak. 
        
        \item Extract the original values of the peak parameter sets \{$\tilde{c}_j$\} at the peak position, again using Eq.~\ref{equation:extract_info}.
        
        \item Calculate the ensemble mean field according to $\bar{f}_{\rm es}(\mathbf{x}) = \xi_i(\mathbf{x}) \xi^{-1}_{ij} (c_j - \tilde{c}_j)$, and obtain the constrained density field via 
        $f(\mathbf{x})$ = $\tilde{f} (\mathbf{x})$ + $\bar{f}_{\rm es}(\bf x)$.
        
    \end{enumerate}
\end{enumerate}

In Figure~\ref{fig:Constrained-IC}, we give a detailed illustration of the above procedure to show how we impose the set of constraints to a random realization of IC. 
All the panels in Figure~\ref{fig:Constrained-IC} show the density field $\delta_x = \rho / \bar{\rho} - 1$ in the IC, with a volume of 20 $\hmpc$ per side (with periodic boundary conditions).
The density fields are extrapolated from $z=99$ (the redshift of IC) to $z=0$.
The leftmost column (two panels) shows the density field of a random, unconstrained Gaussian realization $\tilde{f}(\mathbf{x})$, and the corresponding smoothed $\tilde{f}(\mathbf{x})$ field convolved with a Gaussian kernel of $\rm R_G = 1$ $\hmpc$ (top and bottom panel respectively). 
The black cross marks the peak location of the smoothed density field. 
This is where we choose to impose constraints. 
We apply the peak parameter set extracted from the \textsc{BlueTides} IC for \textsc{BH-BigBH} and \textsc{BT-BigHalo} (c.f. Table~\ref{tab:table1}) to the unconstrained field $\tilde{f}(\mathbf{x})$ by adding the corresponding ensemble mean field constructed via $\bar{f}_{\rm es}(\mathbf{x}) = \xi_i(\mathbf{x}) \xi^{-1}_{ij} (c_j - \tilde{c}_j)$ (c.f. Eq.~\ref{equation:construct_fx}).
The central two panels in Figure~\ref{fig:Constrained-IC} show the final constrained density fields for the two regions \textsc{BT-BigBH} (middle) and \textsc{BT-BigHalo} (right).
The top two panels show the corresponding added ensemble mean fields, which are equivalently the residuals between the constrained and unconstrained density fields. 
The bottom panels give the Gaussian smoothed constrained density field for the two regions to better illustrate the density distribution on large scale.

\begin{table}
\centering
\begin{tabular}{llllllllll}
    \hline
     &$\nu$ & $\rm x_d$ & $a_{12}^2$ & $a_{13}^2$  & $\epsilon$ & $\omega$ \\
     &[$\sigma_0$] & [$\sigma_2$] &  & & [$\kmsmpc$] & [rad]\\
     \hline
    uncons & 3.6 & 3.2 & 2.7 & 3.8 & 40 & 5.6 \\
    \textsc{BT-BigBH} & 4.8 & 5.7 & 1.4 & 2.0 & 34 & 3.9 \\
    \textsc{BT-BigHalo} & 6.0 & 4.5 & 2.0 & 3.0 & 60 & 4.5 \\
    \hline
\end{tabular}
\caption{A subset of 6 peak parameters for \textsc{BT-BigBH} and \textsc{BT-BigHalo} extracted from the initial conditions of the \textsc{BlueTides} simulation with Gaussian kernel of size $\rm R_G = 1$ $\hmpc$.
The corresponding density fields (including the unconstrained random realizations) are shown in Figure~\ref{fig:Constrained-IC}.}
\label{tab:table1}
\end{table}

Table~\ref{tab:table1} lists a subset of 6 out of the 18 peak parameters for the two regions (as well as the peak parameters extracted from the original unconstrained IC).
For example, the density field \textsc{BT-BigBH} has a peak height of $\nu = 4.8\sigma_0$, subject to a tidal field with magnitude  $\epsilon = 34 \kmsmpc$, with the elongation direction lying in the $z$ direction. 
The density field for \textsc{BT-BigHalo} has a height of $\nu = 6 \sigma_0$. The peak is relatively more elliptical compared to \textsc{BT-BigBH}, and is subject to a tidal field with a magnitude of $\epsilon = 60 \kmsmpc$, elongated along the $x$ direction.
Both of these regions are shown evolved to $z=7$ in Figure~\ref{fig:density-field-BH0BH3}.

\subsection{Two peaks: constrained simulations and
\textsc{Bluetides} regions}

With the ICs constructed as described above (and shown in Figure~\ref{fig:Constrained-IC}), we run all the simulations down to $z=7$ and compare their evolved density fields with the regions in the original \textsc{BlueTides} simulation.

Figure~\ref{fig:density-field-BH0BH3} shows the gas density field (colour-coded by temperature) in a region of 10 $\hmpc$ centred on the most massive BH/halo for the unconstrained and constrained simulations of \textsc{BT-BigBH, BT-BigHalo} (top panels). 
For comparison, the bottom two panels below show the actual gas density field of the two sub-regions (of 10 $\hmpc$) directly extracted from the \textsc{BlueTides} simulation at $z=7$.
In addition, the bottom left panel of Figure~\ref{fig:density-field-BH0BH3} also shows the evolution of halo mass ($M_{\rm h}$), stellar mass ($M_{*}$), and BH mass for all of the simulations (constrained runs, the actual \textsc{BlueTides} simulation, and the unconstrained run). \yueying{Here $M_{\rm h}$ is calculated as the virial mass of the halo, by identifying a spherical overdensity region with density 200 times the critical density.}

The comparison of the density field in constrained small volumes to the respective regions of the original large volume \textsc{BlueTides} simulation (as shown in Figure~\ref{fig:density-field-BH0BH3}) illustrates how the CR technique can effectively sculpt the matter distribution around the density peak and resemble the large scale characteristic structures.
With our procedure for generating constrained random fields with peaks specified by the 18 physical characteristics, the density field can be emulated to induce the net gravitational and tidal forces and give rise to a similar morphology to that we see in the large \textsc{BlueTides} simulation.
For example, the large scale structure of the density peak of \textsc{BT-BigBH} is compact with thin filaments that converge radially onto the central halo (as also expected from a relatively low tidal field along the $z$ direction), just as in that region of the \textsc{BlueTides} simulation. 
On the other hand, \textsc{BT-BigHalo} shows a dominant large and thick filament in the density field with a clear elongation in $x$ direction, as expected from the large tidal field lying in that direction.

In our unconstrained simulation, the highest density region has a peak height of $\nu = 3.5 \sigma_0$ (as shown in Table~\ref{tab:table1}). 
In this case, the final halo mass at $z=7$ just exceeds $10^{11} M_{\odot}$, and the central BH mass reaches $10^{6} M_{\odot}$, only slightly above the BH seed mass (note, however, that in this halo, given our BH seeding prescription the BH is also seeded late).
Applying the constraints to this initial density field with the peak parameter set of \textsc{BT-BigBH} in Table~\ref{tab:table1}, the resulting halo grows to $M_{\rm halo} \sim 10^{12} M_{\odot}$, commensurate with the one in the \textsc{BlueTides} simulation (dashed blue line). 
Remarkably, the growth of the BH in constrained simulation also resembles the one from \textsc{BlueTides} simulation, with BH mass exceeds a few $\times 10^{8} M_{\odot}$ by $z=8$.
This is the halo with the earliest and most massive BH in the large volume of BT as discussed in \citet{DiMatteo2017, Ni2018}. 
Finally, the growth of the most massive halo in \textsc{BlueTides} (which does not host the earliest massive BH) is also well recovered in the \textsc{BT-BigHalo} constrained run (orange lines in Figure~\ref{fig:density-field-BH0BH3}). 
In this region, the halo mass approaches $10^{13} M_{\odot}$, while the BH mass has a delayed growth: it is still an order of magnitude below that of the \textsc{BT-BigBH} region at $z=8$, but after a major merger does reach a mass close to the most massive object.

For now, we have shown that the constrained simulation can well reproduce the large scale features of certain density fields from a full hydrodynamical volume, as well as resemble the growth history of the galaxies and BHs in that region.
However, we stress that the growth histories of the galaxies and BHs are not expected to be completely identical to the result of the \textsc{BlueTides} simulation. 
The CR technique constructs one realization of density field conditioned on the desired large scale properties. It is therefore different from a zoom-in technique that re-simulates a certain region selected from a large-volume lower-resolution simulation.

More importantly, our constrained simulations confirm the findings from the large volume \textsc{BlueTides} simulation: the most massive halo with a higher initial density peak does not necessarily contain the largest, earliest BHs. 
This implies that other physical properties of the initial density peak, such as the tidal field and the peak compactness, also play a crucial role in setting the conditions conducive for early BH growth, as we will further discuss in the next sections.



\begin{table}
\centering
\makebox[1\columnwidth][c]{
\begin{tabular}{cccccc}
\toprule
\multicolumn{1}{c}{$f_{\rm 0,G}$} & \multicolumn{3}{c}{$f_{\rm 2,G}$} & \multicolumn{2}{c}{$T_{\rm G}$} \\
\cmidrule(lr){1-1}\cmidrule(lr){2-4}\cmidrule(lr){5-6}
$\nu$ [$\sigma_0$] & $\rm x_d$ [$\sigma_2$] & $\rm a^2_{12}$ & $\rm a^2_{13}$  & $\epsilon$ [$\kmsmpc$] & $\omega$  \\\midrule
5 & 2.2 ($-2\sigma$) & 1.43  & 2.13  & 8 ($-3\sigma$)  & 1.5$\pi$  \\
  & 3.6 (ave)        & 1.33 & 1.77  & 15 ($-2\sigma$) & 1.5$\pi$ \\
  & 4.3 ($+1\sigma$) & 1.29 & 1.66 & 34 (ave) & 1.5$\pi$ \\
  & 5.0 ($+2\sigma$) & 1.25 & 1.56 & 58 ($+2\sigma$) & 1.5$\pi$ \\
  & 5.8 ($+3\sigma$) & 1.22 & 1.49 &    \\
\bottomrule
\end{tabular}
}
\caption{List of the peak parameters we explored in our constrained simulation set. \yy{For Set (i), we explore the role of the peak compactness $x_{\rm d}$ by constraining the $f_{2,\rm G}$ field. For Set (ii), we constrain the tidal field $T_{\rm G}$ by exploring different shear magnitude $\epsilon$. See text for details.}}
\label{tab:paramlist}
\end{table}


\begin{figure*}
\begin{center}
\includegraphics[width=1.9\columnwidth]{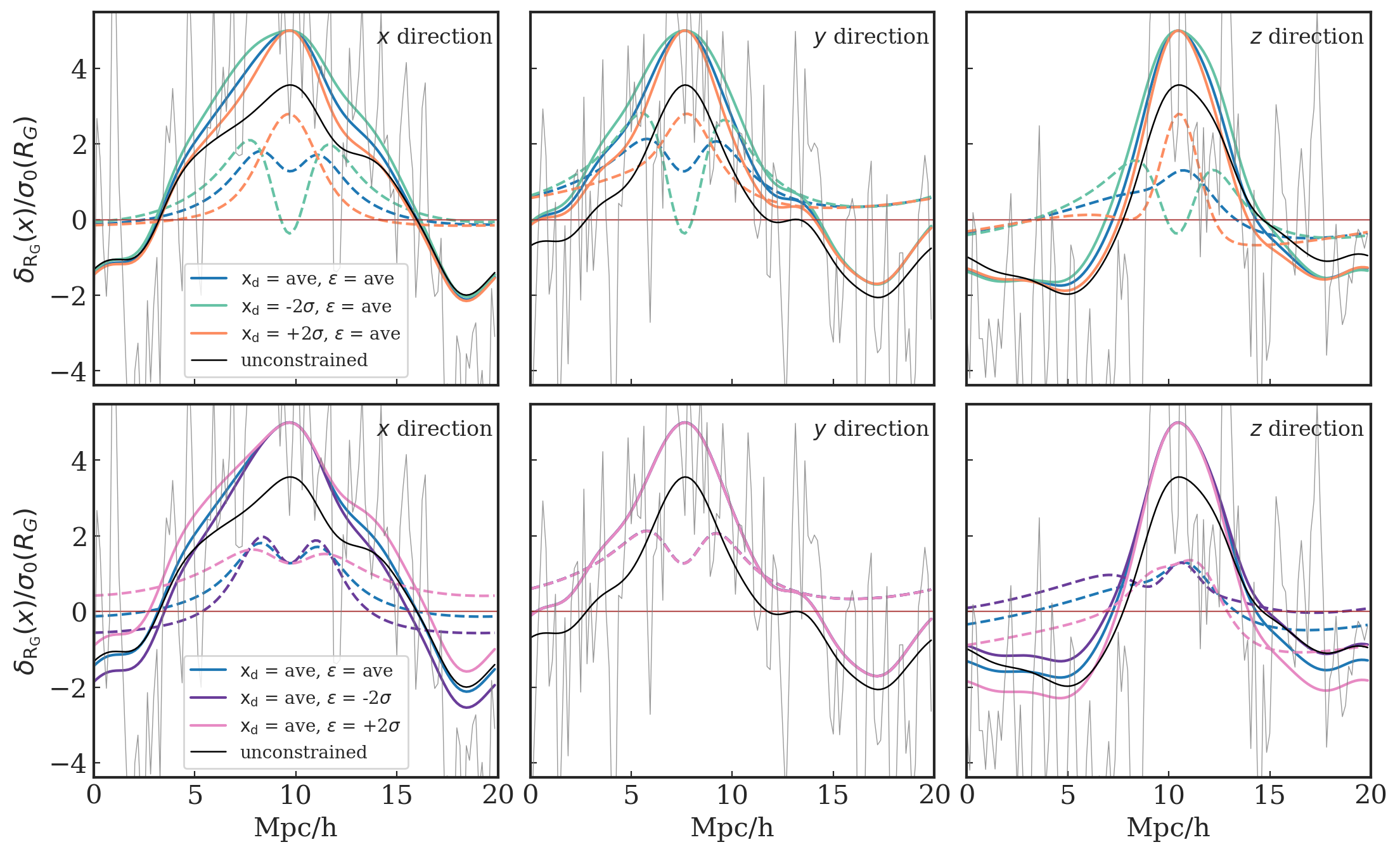}
\end{center}
    \caption{The profiles of the density contrast field in the $x$, $y$ and $z$ directions (from left to right column) for a set of the constrained ICs. The density \yy{contrast} field is linearly extrapolated to $z=0$ and smoothed with Gaussian kernel of width $\rm R_G = 1$ $\hmpc$, and is given in units of $\sigma_0(\rm R_G)$. 
    The thin grey lines in the background of each panels show the unsmoothed original (unconstrained) density \yy{contrast} field $\tilde{f}(\bf x)$. The black solid lines give the smoothed density profiles of $\tilde{f}(\bf x)$. 
    The dashed lines represent the ensemble mean fields added to the unconstrained field that result in the constrained fields shown by the solid coloured lines.
    The top panels show the constrained peaks with different compactness, with green, blue and orange lines corresponding to $\rm x_d = -2\sigma$, $\rm x_d$ = ave, and $\rm x_d = +2 \sigma$.
    The bottom panels show the constrained peaks with varying tidal field magnitudes, with pink, blue and purple lines representing $\epsilon = +2\sigma$, $\epsilon =$ ave and $\epsilon = -2\sigma$.
    Note that the shear angle of the tidal field $\omega = 1.5 \pi$ so that the peak is elongated in the $x$ direction and compressed in the $z$ direction.}
\label{fig:large-profile-IC}
\end{figure*}

\begin{figure}
\begin{center}
\includegraphics[width=1.0\columnwidth]{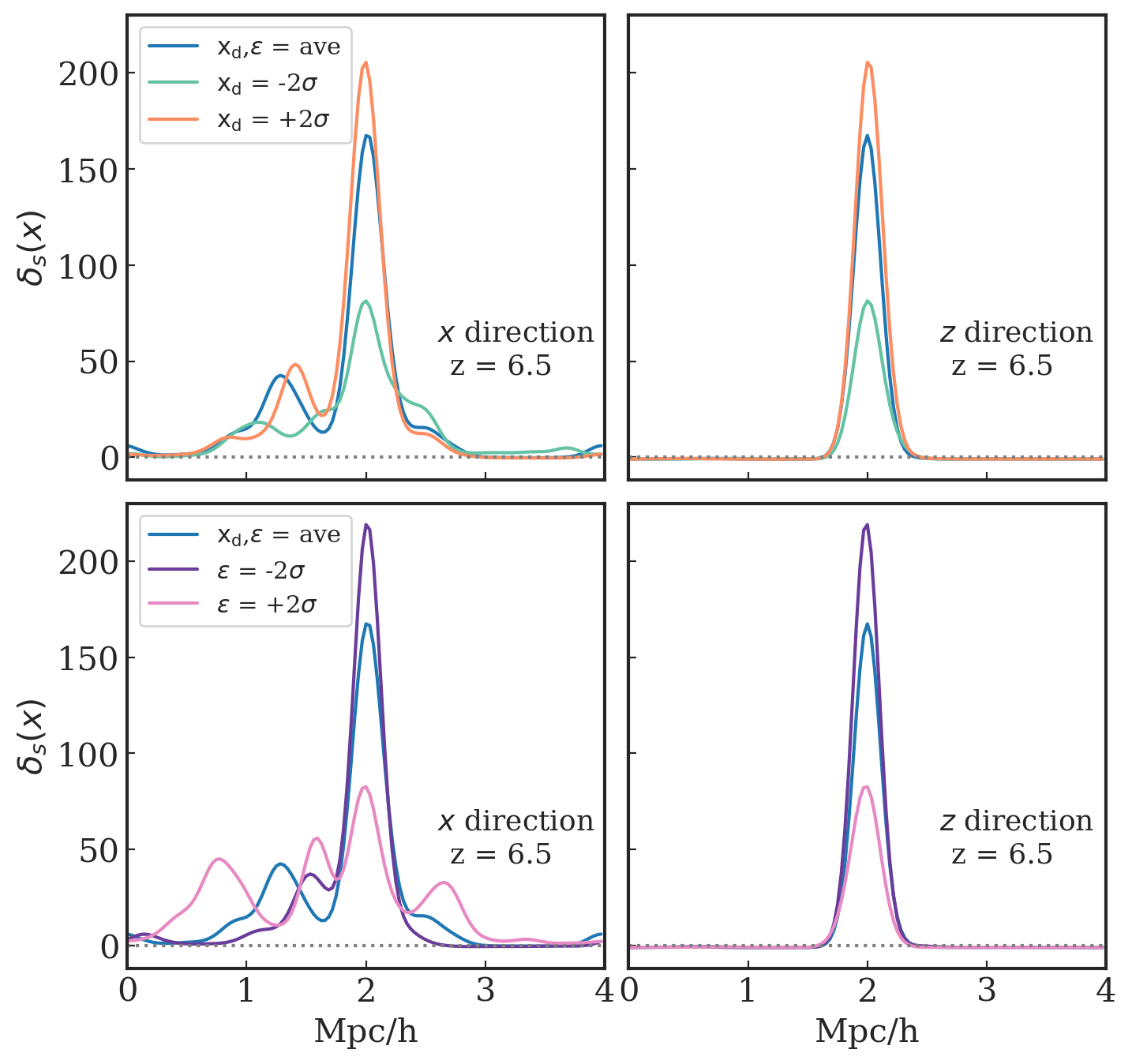}
\end{center}
    \caption{Illustration of the \yy{density contrast} field of constrained simulations at $z = 6.5$ to show the effect of compactness and tidal field. The density contrast field is smoothed over the Gaussian kernel of $\rm R_G$ = 0.1 $\hmpc$ to show the small scale density clumps. The left column gives the profiles in $x$ direction and the right column gives the profile in $z$ direction.
    The top panels show the density profiles from constrained simulation with $\rm x_d = -2\sigma$ (green), $\rm x_d$ = ave (blue), and $\rm x_d = +2 \sigma$ (orange) separately. The bottom panels show the results with averaged compactness and different tidal field magnitude of $\epsilon = -2\sigma$ and $\epsilon = -2\sigma$ in pink and purple lines.
    }
    \label{fig:large-scale-profile-4MPC-RG01}
\end{figure}

\begin{figure*}
\begin{center}
\includegraphics[width=2.1\columnwidth]{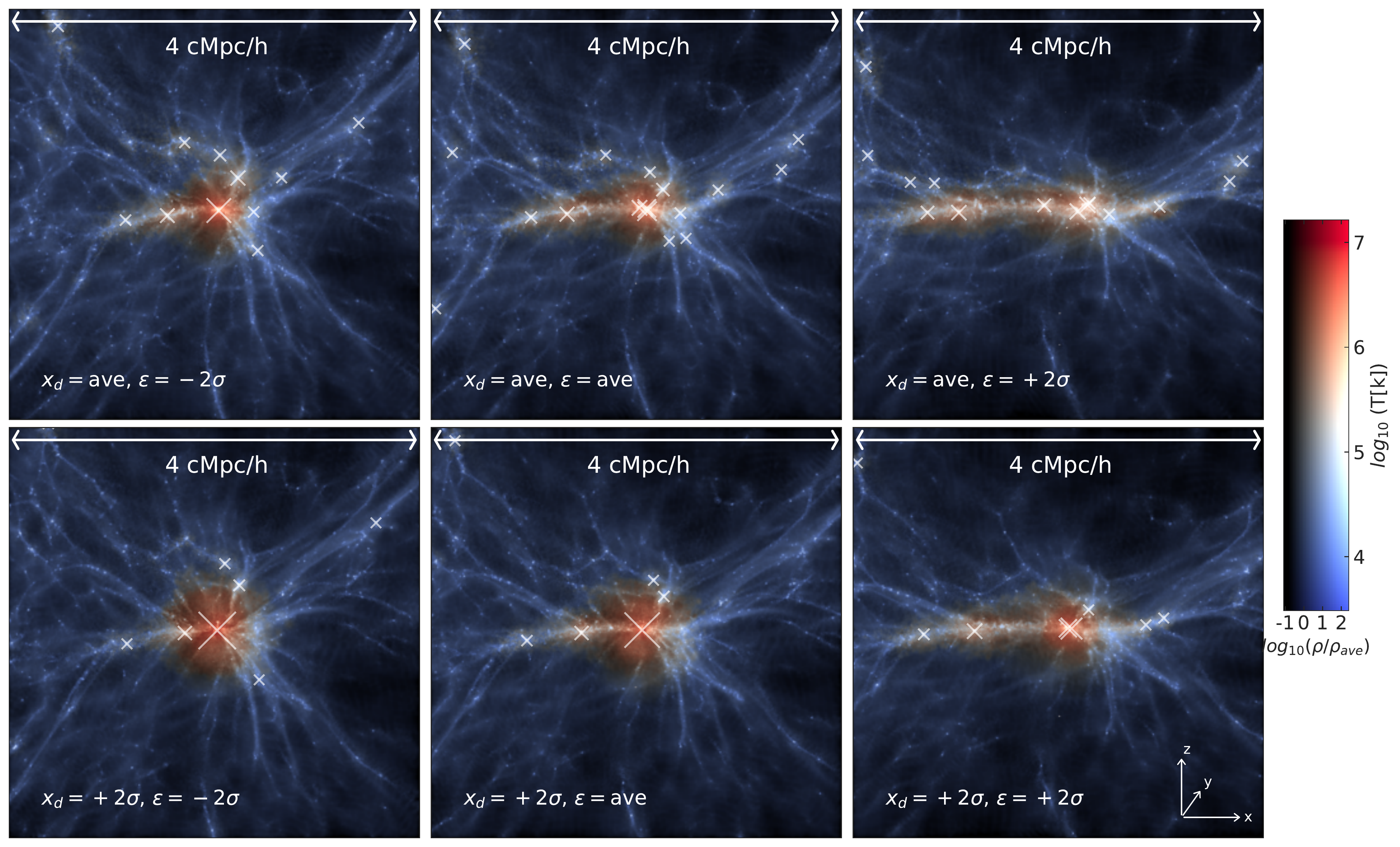}
\end{center}
    \caption{Illustration of the gas density field around the SMBH at $z=6.5$ projected onto the $xz$ plane (with $x$ axis in the horizontal direction and $z$ axis in the vertical direction).
    \yn{For high density regions, the field is colour-coded by the averaged gas temperature, from red to blue indicating warm to cold, as shown by the colour bar aside.}
    Each panel is centred at the maximum of the density peak, showing a zoomed-in region of 4 $\hmpc$ $\times$ 4 $\hmpc$ $\times$ 4 $\hmpc$.
    The white crosses in each panel mark the positions of the BHs, \yy{with the size scaled by the BH mass.}
    The top panels show the constrained simulations with averaged compactness $\rm x_d = 3.6 \sigma_2$ and differing tidal field magnitude of $\epsilon = $ 15, 34 and 58 $\kmsmpc$ from left to right respectively.
    The bottom panels show the corresponding constrained simulations with $+2\sigma$ compactness of $\rm x_d = 5.0 \sigma_2$ with tidal fields ordered the same as above.
    Note that the shear angle of the tidal field $\omega = 1.5 \pi$ so that the peak is elongated in the $x$ direction and compressed in the $z$ direction.}
\label{fig:host-image}
\end{figure*}

\section{Probing the effect of peak parameters on the first quasars}
\label{section4:result}

In this section, we study the impact of large scale features of density peaks on the early BH growth. 
In particular, we focus on density peaks with a height of $\nu = 5 \sigma_0$ and study the impact of the various peak features (as introduced in Section~\ref{subsection:2.2}) on the early quasar growth. 
The structure of this section is as follows.
We first give the parameter distribution of the $5 \sigma_0$ peaks in Section~\ref{subsection4.1:paramspace}.
Section~\ref{subsection4.2:setup} introduces the sets of constrained simulations we carry out in this study.
Section~\ref{subsection4.3:consIC} illustrates the features of those constrained peaks in the initial density fields.
Finally, in Section~\ref{subsection4.4:results}, we give the results of our constrained simulations and show how the initial density peak features affect the early BH growth.

\subsection{Probability distributions of the peak parameters}
\label{subsection4.1:paramspace}

Before investigating which initial density peak features are crucial to the early BH growth, we first need to probe the statistical distribution of the peak parameter values.
In particular, we focus on the parameter distribution for $\nu = 5 \sigma_0$ density peaks on a scale of $\rm R_G = 1$ $\hmpc$ in our 20 $\hmpc$ simulation box.
To this end, we apply a bootstrapping method as the easiest approach to extract the distribution. 
We first generate a large set of \yueying{different} random realizations of ICs on which we impose the $5 \sigma_0$ peak constraint on a random position. 
Then we read out the resultant density peak parameters of those $5 \sigma_0$ peaks to get the overall distribution.

Figure~\ref{fig:param-distribution} shows the result for the probability distributions of all the peak parameters conditioned on $\nu = 5 \sigma_0$, drawn from $10^5$ \yueying{different} constrained ICs of random realizations.  
In particular, the top left panel shows the distribution of compactness, $\rm x_d$ (in units of $\sigma_2 (\rm R_G)$).
The probability distribution of $\rm x_d$ only depends on the peak height, with higher density peaks having a larger $\rm x_d$ \citep{Bardeen1986}. 
For density peaks with $\nu = 5 \sigma_0$, we can see that the distribution of $\rm x_d$ centres at $\rm x_d = 3.6$, with $\pm 2\sigma$ located at $\rm x_d = 5.0$ and 2.1 correspondingly, as shown by the vertical dotted lines.

The top middle panel gives the 2D probability distribution of the ellipticity of the density peak $P(a_{12}^2, a_{13}^2)$, with $a_{12}^2$, $a_{13}^2$ introduced in Section~\ref{subsection:2.2}.
Note that the ellipticity distribution is only dependent on the peak compactness $\rm x_d$, with higher values of $\rm x_d$ driving lower ellipticity. 
The blue data points plot the 2D histogram of $a_{12}^2$, $a_{13}^2$ for the constrained density peaks with $\rm x_d \sim 3.6$. The \yy{blue solid lines give} the contour of the probability distribution $P(a_{12}^2, a_{13}^2 | \rm x_d = 3.6)$ with confidence level of 97.5\% and 90\% from the theoretical estimation by \cite{Bardeen1986}.
The orange and red dashed lines give the 97.5\% contours of $a_{12}^2$, $a_{13}^2$ distribution conditioned on $\rm x_d$ = 2.1 and $\rm x_d$ = 5.0 separately, with the cross marking out the point with the maximum probability.
We \yy{show} that the distributions of $\rm x_d$ and $a_{12}$, $a_{13}$ drawn from our constrained realizations are consistent with the theoretical predictions of \cite{Bardeen1986}, demonstrating that the density field generated by the CR technique is a properly sampled realization of the entire probability distribution of the Gaussian random field.

The top right panel of Figure~\ref{fig:param-distribution} gives the
distribution of the peculiar velocity of the peak in three directions $v_x$ (blue colour), $v_y$ (green colour) and $v_z$ (red colour) in unit of $\kms$.
We note that the variance of the peak velocity drawn from our constrained realization is about 60 $\kms$, much smaller than the analytical estimate of the peculiar velocity variance $\sigma_v(\mathrm{R_G}) = H_0 F_{\Omega} \sigma_{-1}(\mathrm{R_G}) \sim 330 \kms$.
This is expected because the peculiar velocity is induced by the mass density fluctuation on large scales, (note that $\sigma_{-1}$ is weighted by $k^{-1}$).
The variation of $\sigma_{-1}$ is mostly contributed by scales larger than our box size of 20 $\hmpc$ (see Appendix~\ref{Appendix A} for more details).
Hence, the simulation box size limits the peculiar velocity we can generate, in such a way that it is unlikely to have a large peculiar velocity field in our constrained simulations.

The bottom left panel of Figure~\ref{fig:param-distribution} shows the distribution of the tidal field magnitude (shear scalar) $\epsilon$ in the unit of $\kmsmpc$.
The distribution of the tidal tensor is not significantly affected by the box size, since the variance of the tidal field is related to $\sigma_0$, mostly contributed by scales within the box.\footnote{see Appendix A for further discussion on the effects of box size}
The bottom middle panel gives the distribution of the shear angle $\omega$: the relative strength of the tidal field along the three principal axes. It peaks at $\omega = 1.5 \pi$, corresponding to the case when the three eigenvalues of the tidal tensor have the sign of (-,0,+), as illustrated in the 4th panel of Figure~\ref{fig:contour3D}.

Apart from the parameters discussed above, there are six Euler angles that specify the direction of the mass ellipsoid and the tidal field.
The tidal field tends to align itself along the principal axes of the mass tensor of the density peak. It tends to elongate along the long axis of the mass ellipsoid and contract with respect to the shortest axis of the ellipsoid.
To illustrate this effect, we plot in the bottom right panel of
Figure~\ref{fig:param-distribution} the intersection angles between the direction of the tidal tensor and the mass ellipsoid.
The green colour shows the distribution of the intersection angle between the major axis of the mass ellipsoid and the direction of the compression due to the tidal field (i.e., the eigenvector of the tidal tensor with the smallest, negative eigenvalue).
The blue-coloured distribution instead shows the intersection angle between the major axis of the mass ellipsoid and the direction of the tidal field, (i.e., the eigenvector of the tidal tensor with the largest, positive eigenvalue). 
As one might expect, the \yy{green} distribution peaks at $0.5\pi$ while the blue distribution peaks at $0$ and $\pi$, indicating that the major axis of the mass ellipsoid tends to be perpendicular to the compression direction of the tidal field, and align with the elongation direction.

Given the probability distribution for all the constraints associated with the peak parameters, we can get hints of which of the peak features are more relevant to the growth of the first SMBH in case of \textsc{BT-BigBH} and form the most massive halo in \textsc{BT-BigHalo}.
Comparing the distributions in Figure~\ref{fig:param-distribution} to the values in Table~\ref{tab:table1}, we find that the primordial peak of \textsc{BT-BigBH} is extremely compact, with $x_{\rm d}$ nearly  $+3\sigma$ above the mean value of the overall distribution.
Meanwhile, in the \textsc{BT-BigHalo} case whose primordial density peak has a rather extreme height of $\nu = 6 \sigma_0$, the corresponding SMBH does not grow as rapidly as for \textsc{BT-BigBH}.
We find that the peak of \textsc{BT-BigHalo} is less compact, with $x_{\rm d}$ slightly above the average of the overall distribution. (Note that $P(x_{\rm d})$ peaks at $x_{\rm d} = 4$ for $\nu = 6 \sigma_0$ peaks.) 
Moreover, it lies in a large tidal field with magnitude $\epsilon = 60 \kmsmpc$, more than $+2 \sigma$ above the mean value of the $\epsilon$ distribution.
We find that peak compactness and tidal field are indeed two of the most important features of the primordial density peak that affect the early growth of SMBHs.
We carry out a more detailed study of these two parameters in the next sections.

\subsection{Design of the constrained simulations}
\label{subsection4.2:setup}

With the detailed probability distributions for the peak parameters for a $\nu = 5 \sigma_0$ peak, and the guidance provided by the comparison of the peak characteristics of the \textsc{BT-BigBH} and \textsc{BT-BigHalo} cases, we now explore systematically the role of (a) peak compactness and (b) tidal field constraints on the growth of an early massive BH. 

Note that for all the constrained simulations in this study, we impose constraints at the same position and is based on the same random unconstrained realization $\tilde{f}(\bf x)$. This allows us to study the effect of various peak parameters on an equal footing.

In Table~\ref{tab:paramlist} we provide a summary of the systematic parameter study we perform varying the compactness and tidal field magnitude separately. 
In set (i) we vary the compactness, starting from mean value of $\rm x_d = 3.6$ and increasing it to $+1 \sigma$, $+2 \sigma$, $+3 \sigma$ of the $\rm x_d$ distribution ($\rm x_d = 4.3, 5.0, 5.8$ respectively). 
As illustrated in the top panels of Figure~\ref{fig:large-profile-IC}, increasing the peak compactness affects the curvature of the peak, making it increasingly narrower in the inner regions. 
As the compactness is the second-order derivative of the density field, this constraint determines the density distribution in the immediate vicinity of the peak.
Moreover, since the probability distribution of ellipticity $a_{12}$ and $a_{13}$ is (only) conditioned on the compactness, for each $\rm x_d$ we apply the $a_{12}$, $a_{13}$ value corresponding to the maximum of the probability distribution. 
Note that a more compact peak tends to be more spherical, with correspondingly lower values of $a_{12}$ and $a_{13}$ (see Table~\ref{tab:paramlist}). 
We also run a realization with a  $-2 \sigma$ value of the compactness to contrast the results of a high curvature peak with one of low curvature.

For (ii) the tidal field constraint, we explore different shear magnitudes $\epsilon =  8, 15, 34, 58 \kmsmpc$, which correspond to $-3 \sigma$, $-2 \sigma$, the average value, and $+2 \sigma$ respectively. 
These values are based on the  $\epsilon$ distribution shown in Figure~\ref{fig:param-distribution}.
We fix the shear angle to be $\omega = 1.5 \pi$ for all of our constrained simulations so that the tidal field is elongated in one direction and equally compressed in another direction, as illustrated in the third panel of Figure~\ref{fig:contour3D}. 

To give a clear illustration of the field direction, for all of our constrained simulations, we orient the long axis of the peak ellipsoid along the $x$ direction, and the shortest axis along the $z$ direction. 
We align the tidal field correspondingly, directing the elongation of the tidal field along the $x$ axis, and compression direction along the $z$ axis, to account for the fact that the tidal field tends to align with the peak mass ellipsoid as shown in the last panel of Figure~\ref{fig:param-distribution}. 

Finally, for the peculiar velocity, our preliminary test shows that it has a negligible impact on early BH growth. It is expected since the peculiar velocity is induced by the asymmetry of the larger scale matter distribution. 
Therefore, we just set $v_x = v_y = v_z = 0 \kms$ for all the constrained density peaks, which means that there is no gravitational acceleration on scale of $\rm R_G = 1$ $\hmpc$ at the peak position.

\subsection{The effects of peak compactness and tidal field in constrained ICs}
\label{subsection4.3:consIC}

We first illustrate the effect of constraining the compactness and tidal field of a density peak in the ICs.
In Figure~\ref{fig:large-profile-IC}, we plot the density profile across the peak maximum in the $x$, $y$ and $z$ directions for a set of constrained peaks with various $\rm x_d$ and $\epsilon$ values as described in Section~\ref{subsection4.2:setup}.
For illustration, the density contrast field is linearly extrapolated to $z=0$ and is smoothed with a Gaussian kernel of width $\rm R_G = 1$ $\hmpc$. 
The $y$ axis gives the $\delta (x)$ in units of $\sigma_0(\rm R_G)$ (c.f. Eq.~\ref{equation:sigma0}). 
For comparison, the black lines in each panel show the density profile of the original unconstrained realization. 
The dashed lines represent the ensemble mean fields $\bar{f}_{\rm es}(\mathbf{x}) = \xi_i(\mathbf{x}) \xi^{-1}_{ij} (c_j - \tilde{c}_j)$ added to the unconstrained field (c.f. Eq.~\ref{equation:construct_fx}). 
And the solid coloured lines are the corresponding profiles of the constrained density contrast fields.
The top panels show the effects on the peaks when different values of peak compactness are set. 
The bottom three panels show the constrained peaks with varying tidal field magnitudes.
As the angle of the tidal field $\omega = 1.5 \pi$, the peak is elongated in the $x$ direction, compressed in the $z$ direction, \yy{and has no difference in the $y$ direction.}

By construction, all of the constrained peaks have a height of $5 \sigma_0$. 
Figure~\ref{fig:large-profile-IC} illustrates that a peak with low compactness ($\rm x_d = -2 \sigma$) is more extended (in the innermost regions) than a peak with larger compactness.
The added ensemble mean field (green dashed line) reshapes the matter distribution of the original density field and makes the resultant smoothed density profile flatter at the maximum.
In contrast, in the bottom panels, the added ensemble mean field corresponding to different tidal field magnitudes is not as dramatically different in the inner regions. 
Increasing the tidal field sculpts the matter distribution on larger scales.
As expected, the constrained field with a larger tidal field is more extended in the wings of the profile (in our case, by construction, only in the $x$ direction).
Next, we investigate the consequences of these different physical parameters in the ICs on the later growth on the halo and in particular of the central BHs.


\begin{figure}
\begin{center}
\includegraphics[width=1.0\columnwidth]{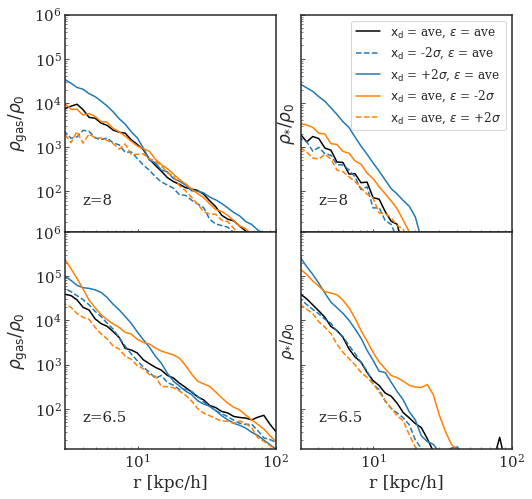}
\end{center}
    \caption{
    The gas density profiles (left column) and stellar density profiles (right column) around the most massive BH in different constrained simulations at $z = 8.0$ (top panels) and $z = 6.5$ (bottom panels). 
    The $x$ axis gives the comoving distance to the central BH.
    The gas and stellar density is calculated by averaging over the spherical shells around the BH and is given in the unit of $\rho_0$, which is the averaged matter density of the universe.
    The black line gives the result from the constrained simulation with averaged peak compactness $\rm x_d = 3.6 $ and tidal field magnitude $\epsilon = 34 \kmsmpc$. 
    The blue solid and dashed line represent the one with averaged tidal field and compactness $\rm x_d = 5.0 $ (+2$\sigma$) and $\rm x_d = 2.2 $ (-2$\sigma$) respectively.  
    While the orange solid and dashed line represents the constrained simulation with averaged compactness $\rm x_d$ and tidal field magnitude $\epsilon = 15 \kmsmpc$ (-2$\sigma$) and  $\epsilon = 58 \kmsmpc$ (+2$\sigma$) respectively.}
    \label{fig:density-profile}
\end{figure}

\begin{figure}
\begin{center}
\includegraphics[width=1.0\columnwidth]{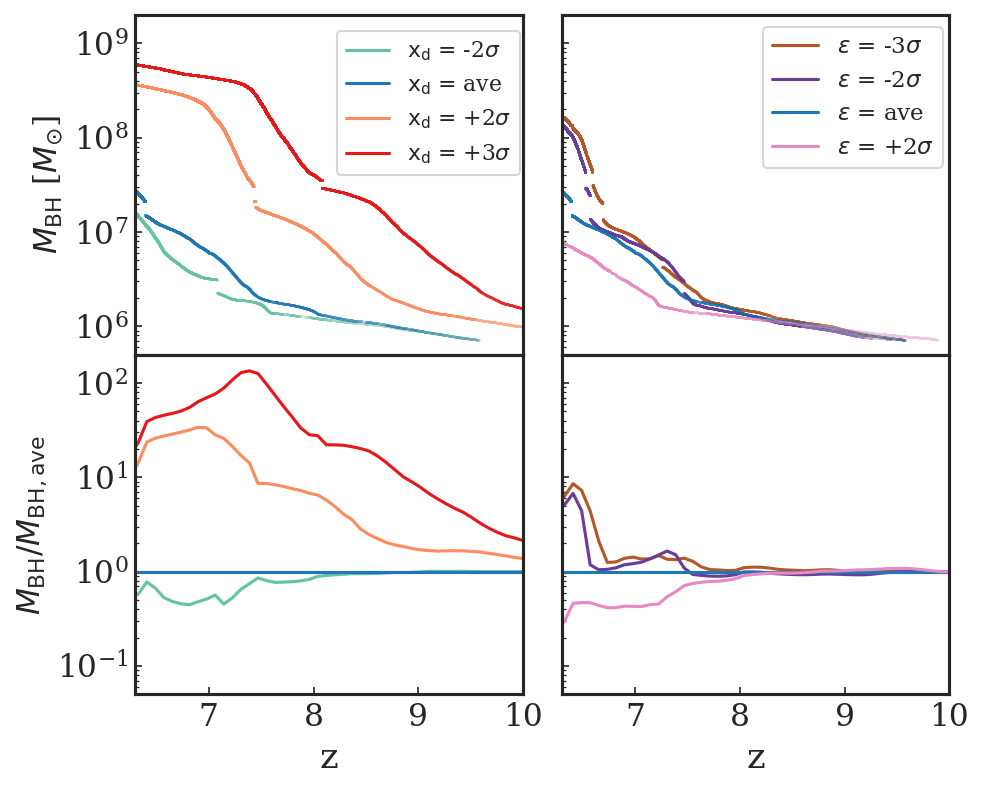}
\end{center}
    \caption{BH growth history in constrained simulations with varying peak parameters. The blue line represent the run with averaged peak compactness of $\rm x_d = 3.6 \sigma_2$ and tidal field magnitude $\epsilon = 34 \kmsmpc$. 
    The green, orange and red lines in the left column are the constrained run with different peak compactness of $\rm x_d = 2.1$ (-2$\sigma$), $\rm x_d = 5.0$ (+2$\sigma$) and $\rm x_d = 5.8$ (+3$\sigma$) respectively. The pink, purple and brown lines in the right column represent the BH growth with varying tidal field magnitude of $\epsilon = 58 \kmsmpc$ (+2$\sigma$),  $\epsilon = 15 \kmsmpc$ (-2$\sigma$) and  $\epsilon = 8 \kmsmpc$ (-3$\sigma$) respectively. 
    The top panels give the growth history of BH mass as a function of redshift. 
    The gaps in the plot correspond to the BH merger event. 
    We only trace the more massive progenitor for each merger event.
    The bottom panels give the ratio of $M_{\rm BH}$ compared with the \yy{one residing in the averaged peak compactness and tidal field} (the blue line).}
    \label{fig:BH-growth-xd-epsilon}
\end{figure}

\begin{figure} 
\includegraphics[width=1.0\columnwidth]{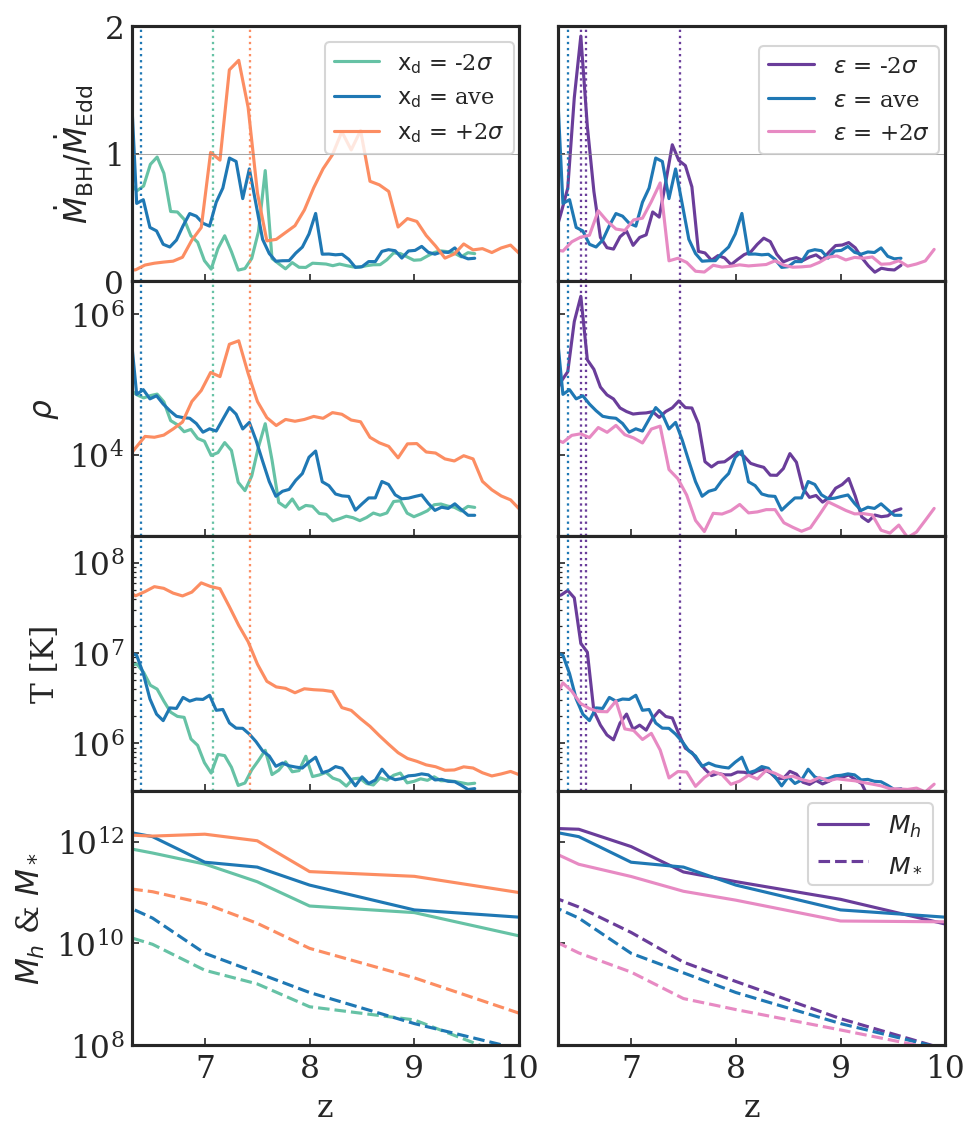}
\caption{The detailed evolution history of the BHs in constrained simulations with varying peak compactness (left column) and tidal field (right column). The colour convention is the same as Figure~\ref{fig:BH-growth-xd-epsilon}. 
\textit{Top panel:} The Eddington accretion ratio $\dot{M}_{\rm BH}/M_{\rm Edd}$ of the BH growth. 
\textit{Second panel:} The averaged gas density in the vicinity of the BH, in the unit of the averaged matter density of the universe $\rho_0$. 
\textit{Third panel:} The gas temperature in the BH surrounding. Both gas density and temperature are averaged over the gas properties within the SPH smoothing kernel of the BH. 
\yueying{\textit{Bottom panel:} The virial mass of the host halo $M_{\mathrm h} = M_{200}$ in solid line and the stellar mass ($M_*$) within the virial radius in dashed line. The vertical dotted lines in each panel marks the time of the BH merger event.}
}
\label{fig:BH-details-xd-epsilon}
\end{figure}

\subsection{Results of the constrained simulations}
\label{subsection4.4:results}
In this section, we show the results of the constrained simulations probing different parts of the parameter space of the constraints. 
In particular, we look at the impact of the different peak constraints on the early growth and evolution of the most massive BHs at $z>6$.

\subsubsection{Evolution of the density field}

We run all the constrained simulations down to $z = 6.5$.
We first look at the evolved density fields in the vicinity of the SMBH residing in the centre of the density peak. 
Figure~\ref{fig:large-scale-profile-4MPC-RG01} shows the 1D projections of the peak profiles in the $x$ and $z$ directions within 4 $\hmpc$ around the central SMBH at $z = 6.5$ for different constrained simulations. 
The density contrast field shown here is smoothed with a Gaussian kernel of width $\rm R_G$ = 0.1 $\hmpc$ to illustrate the density clumps on relatively small scales (of 100 $\hkpc$).

The top panels show density profiles from constrained simulations with realizations of different compactness, $\rm x_d = -2\sigma$, $\rm x_d = $ ave, and $\rm x_d = +2 \sigma$.
It is evident that at $z=6.5$ the more compact peaks have grown to an even narrower density peak than a mean $5\sigma$ peak. 
A similar result is obtained in the constrained simulations with a low tidal field, as shown by the purple line in the bottom panels, which again reveal an extremely narrow and isolated overdensity compared to realizations with larger tidal stresses.
In the rest of this section, we investigate how the enhanced growth of peaks for high compactness and low tidal field also lead to larger gas inflows and eventually higher accretion rates onto the central SMBH.

Figure~\ref{fig:host-image} shows images of the gas density colour-coded by temperature in a $4 \times 4 \times 4$ $\hmpc$ sub-region surrounding the most massive BH (which is embedded in the middle of the density peaks).
These images illustrate the relatively large scale gas density distributions around the BHs. 
The top and bottom rows show the results from the constrained simulations with compactness $x_{\rm d} = \mathrm{ave}$ and $x_{\rm d} = +2 \sigma$ respectively. 
From left to right in each row we show the results for different tidal fields increasing from $- 2\sigma$, to mean and to $+2\sigma$ respectively.
Each panel in Figure~\ref{fig:host-image} shows the density field projected onto the $xz$ plane. 

Comparing the gas density fields it is evident that more compact, concentrated, initial density peaks (due to high compactness and/or low tidal field) also lead to a much more concentrated gas density environment around the central BHs. 
Around peaks with lower tidal field and large compactness (as shown by the bottom left panel), the gas density is strongly peaked, the surrounding filaments are relatively cold, and accretion occurs in different directions through separate misaligned filaments. 
We will see that under these conditions strong gas infall is common and favours the growth of the SMBH at these early epochs.
In contrast, in realizations with a large tidal field and low compactness (as shown in the top right panel), a significant filament forms on scales larger than the typical size of the halo. 
Gas is accreted from different directions with deceleration along the major filament ($x$ direction) and acceleration (squeezed) along the $z$ direction. 
We discuss the evolution of BH growth and gas environment in more detail in the following sections.

\subsubsection{Gas density profile in the BH host galaxy}

The accretion onto the SMBH is sensitive to the gas environment in its surroundings.
To further explore the effect of initial density peak parameters on the gas inflow rates into the BH host galaxy, we investigate the gas and stellar density profiles in the BH host galaxy.
Figure~\ref{fig:density-profile} shows the averaged gas density profiles (left column) and stellar density profiles (right column) as a function of the comoving distance to the central BH.
The top panels show the profiles at $z = 8.0$, and the bottom panels show the result at $z = 6.5$.

The gas and stellar densities are calculated by averaging over the spherical shells around the BH and are given in the unit of $\rho_0$, the averaged matter density of the universe.
The black line gives the result from the constrained simulation with average peak compactness and tidal field magnitude.
The blue lines represent the cases with average tidal field and compactness $\rm x_d = +2 \sigma$ (solid) and $\rm x_d = -2 \sigma$ (dashed).  
The orange lines on the other hand give the cases with averaged compactness and tidal field $\epsilon$ = -2$\sigma$ (solid) and +2$\sigma$ (dashed).

As shown by the blue solid lines, at both $z = 8$ and $z = 6.5$, the gas and stellar density profiles are much steeper when the initial density peak is more compact. 
For the low tidal field scenario, there is not a large difference in the density profiles at $z = 8$. 
However, the profiles are significantly enhanced by $z = 6.5$. 
The large scale spherical matter distribution brought about by the low tidal field helps the infalling gas to form a cuspy inner gas profile.

Conversely, for the constrained simulations with low compactness or high tidal field, 
the corresponding gas and stellar density profiles are shallower at $z=8$ compared with the others.
Note however that the largest differences in the inner profiles are seen at earlier times $z=8$ and narrow at $z=6.5$. 
As we will discuss in the following section, when the central BHs in these galaxies grow larger, the AGN feedback also starts to play a more important role and interplay with the surrounding gas environment \citep[see also][]{Ni2018,Ni2020}.
Those complex astrophysical processes would modulate the surrounding gas field and narrow down the initial difference at later epochs.

\subsubsection{BH evolution history}

In Figure~\ref{fig:BH-growth-xd-epsilon}, we show the growth history of the BH mass in the same set of the constrained simulations, varying either compactness or tidal field magnitudes. 
In particular, the left column shows the BH mass and accretion rate evolution for  constrained simulations with averaged tidal field and varying peak compactness: 
-2$\sigma$, mean, +2$\sigma$ and +3$\sigma$ (green, blue orange and red lines respectively).
On the right, we show the corresponding BH mass evolution for constrained simulations with mean compactness and a varying tidal field magnitude, with pink, blue, purple and brown lines representing +2$\sigma$, mean,  -2$\sigma$ and -3$\sigma$ respectively. 
The small gaps in the $M_{\rm BH}$ growth correspond to BH merger events. 
Here we trace the more massive progenitor for each merger event.

The BH mass growth in Figure~\ref{fig:BH-growth-xd-epsilon} indicates that a more compact initial 5$\sigma$ peak significantly boosts the BH growth at early redshifts.
The BH residing in the most compact peak (with $x_{\rm d} = +3 \sigma$) has grown to a mass of $M_{\rm BH} = 5 \times 10^8 M_{\odot}$ at $z \sim 7.5$, about 2 orders of magnitude larger than its counterpart residing in average compact $5\sigma$ peak (e.g. shown by the blue line).
Although we note that for the 2$\sigma$ and 3$\sigma$ compactness models the BHs are also seeded increasingly earlier, which could enhance the effect. 
The right columns in Figure~\ref{fig:BH-growth-xd-epsilon} show that BHs residing in low tidal field regions also grow faster compared to BHs embedded in high tidal field regions. 
Large tidal fields can induce a significant delay in the BH growth. 
For example. the BH grown in the $\epsilon = 2 \sigma$ tidal field (pink line) only reaches $<10^7 M_{\odot}$ at $z = 6.5$, order of magnitude smaller than its counterparts residing in an average or lower tidal field.

As clearly shown in the bottom panels of Figure~\ref{fig:BH-growth-xd-epsilon}, the enhanced gas density and cuspy inner profiles induced by a low tidal field and high compactness in a 5$\sigma$ peak can result in enhanced BH mass growth by factors up to 10 or 100 times larger than average.
To investigate more directly the influence of the IC peak parameters on  BH accretion, we show in Figure~\ref{fig:BH-details-xd-epsilon} the BH accretion rates as well as the associated gas densities and temperatures in the vicinities of the BHs.
The top panels show the Eddington accretion ratio $\dot{M}_{\rm BH}/M_{\rm Edd}$.
Note that $M_{\rm Edd}$ is a function of $M_{\rm BH}$ (c.f. Eq.~\ref{equation:Meddington}).
The gas density and temperature are calculated by taking an average over neighbouring gas particles within the SPH smoothing kernel of the BH, which is roughly on scales of 1 $\hkpc$.

The \yueying{second} panel of Figure~\ref{fig:BH-details-xd-epsilon} shows that the gas density is enhanced in the innermost region around the BH as a result of the increased compactness or low tidal field (as also expected from the gas density profiles in Figure~\ref{fig:density-profile}). 
This enhanced surrounding gas density directly leads to a boosted BH accretion rate.
In other words, the physical characteristics of the peak in the ICs play an important role in regulating the gas inflow rates and therefore have a large impact on the BH growth history in the early phases. 

\yueying{The third panel of Figure~\ref{fig:BH-details-xd-epsilon} shows the averaged temperature of the gas within the accretion kernel (SPH smoothing kernel) of BH.}
We see a steep increase in the temperature of the accreting gas as the BH grows larger.
This is brought about by the AGN feedback that dumps part of the BH accretion energy onto its surroundings, heating and clearing out the nearby fueling gas.
This process, in turn, suppresses the gas density and accretion onto the BH itself.
Therefore, with the modulation of surrounding gas brought about by the AGN feedback, the BH growth can not consistently stay in a high accretion mode.
The early growth of a BH in the high compactness and low tidal field simulations will be eventually caught up by a BH with average parameters of a $5\sigma$ peak later on, at $z < 6$.

\subsubsection{Effect of the peak compactness and tidal field}

\begin{figure*} 
\includegraphics[width=1.8\columnwidth]{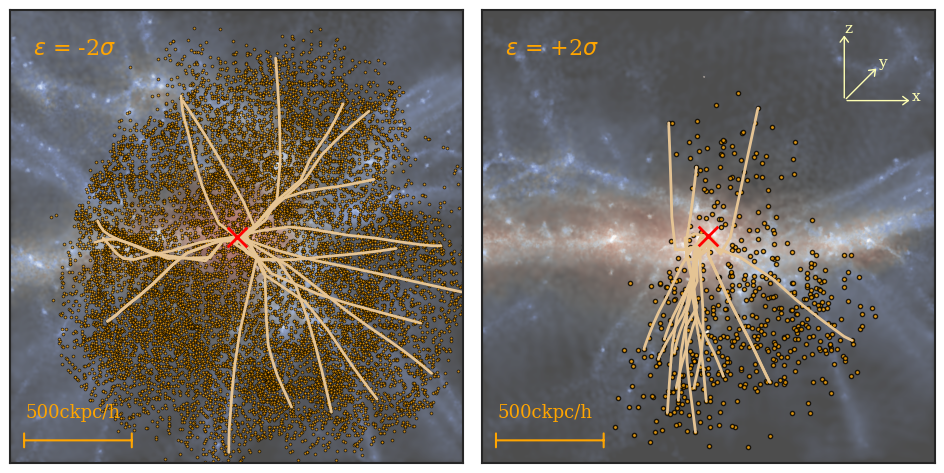}
\caption{Illustration of the gas particle distribution and trajectories that contribute to the accretion onto the central BHs in two constrained simulations with tidal field $\epsilon = -2 \sigma$ (left panel) and $\epsilon = +2 \sigma$ (right panel). 
The background in each panel shows the overall gas density field around the most massive BH, with the central red cross giving the central BH position at $z = 6.5$. For both simulations, we trace the gas particles within 3 $\hkpc$ from the central BH at $z=6.5$ back to $z = 15$ and show their distribution in orange dots. From them, we randomly select 20 gas particles and plot their trajectories from $z=15$ to $z=6.5$ in yellow lines.}
\label{fig:Gas-trajectory}
\end{figure*}

\yueying{
Now we discuss in more detail how the peak compactness and tidal field affects the central BH growth.}

\yueying{
We first look at how different constraint parameters affects the evolution of the BH hosts. 
The bottom panel of Figure~\ref{fig:BH-details-xd-epsilon} shows the growth histories of the host halo $M_{\rm h}$ (solid lines) and host galaxy (dashed lines), where $M_{\rm h}$ is calculated as the virial mass of the halo.
We can see that high peak compactness induces an earlier formation of the halo at high redshift, which also results in earlier seeding of the central BH. 
Halos embedded in different tidal field strengths have a similar mass at $z=10$. However, the halo mass growth in the case of a large tidal field is delayed compared with the one in the low tidal field.
}
\yueying{
Studies of the tidal field effect on the formation of galactic halos have been carried out in some earlier work \citep[e.g.][]{Borzyszkowski2017}, demonstrating that matter cannot effectively accrete onto the halo along the filament. 
This effect accounts for the delay of the mass growth for host halos embedded in a large tidal field region.
}

\yueying{
The growth of the host galaxies in different constrained simulations shows a similar trend as the BH growth. The star formation is sensitive to the gas density in the halo centre. Therefore, the environment that boosts the BH growth also leads to a relatively high stellar mass. We will further discuss the BH and their host galaxy stellar components in Section~\ref{section5:Obs}.
}

Peak compactness and tidal field also affect the BH mergers. 
The vertical dotted lines in Figure~\ref{fig:BH-details-xd-epsilon} marks the time when the BH undergoes a merger event in each respective simulation. 
We can see that the BH merger events are typically followed by a rapid increase in the central density around BH which in turn leads to a boost of the BH accretion.
For the simulation with large compactness or low tidal field, the merger event typically happens earlier. 
In the case of a more compact density peak, e,g., the orange line ($x_{\rm d} = +2\sigma$), the merger happens earlier with the earlier formation of the small parent halos and a more compact spatial distribution of matters.
On the other side, in simulations with large tidal fields (e.g. $\epsilon=+2\sigma$ pink line), BH merger events got delayed, as the filamentary matter distribution lying along the tidal field decelerates the mergers of structures in the $x$ direction.

\yueying{
The BH resides in the innermost central region of the halo, its growth is determined by the local environment of the gas properties within a few ckpc, which is on a small scale.
It is somewhat more intuitive to deduce that the peak compactness affects the BH growth, as we discussed in Section~\ref{subsection:2.2}. 
The compactness modulates the small scale structure of the peak and this can directly affect the BH local environment. 
As shown in the bottom panel of Figure~\ref{fig:BH-details-xd-epsilon}, the high compactness leads to an earlier collapse of the central halo while also inducing a high-density central region fueling the rapid BH growth. 
}

\yueying{
On the other hand, however, the tidal fields instead modulate the matter distribution at larger scales and affect the halo environment. 
It is rather interesting that the tidal field at large scales can still dictate the small-scale environment in the innermost region of the halo and therefore affects the BH growth.
}

\yueying{
In Figure~\ref{fig:Gas-trajectory}, we further demonstrate the effect of large-scale tidal fields on the gas accretion onto the innermost central region around BH in two simulations of high and low tidal field constraints.
The background of Figure~\ref{fig:Gas-trajectory} in the left and right panel show the gas environment around the most massive BH at $z = 6.5$ for constrained simulations with mean compactness and tidal field $\epsilon = -2\sigma$, $\epsilon+2 \sigma$ separately.
To inspect the origin and trajectories of the gas particles that actively participate in the accretion onto the BH, we trace the gas particles within 3 $\hckpc$ from the central BH from $z=6.5$ back to $z=15$. The orange dots in Figure~\ref{fig:Gas-trajectory} plot the spatial distribution of gas particles at $z=15$ that end up in the BH vicinity at $z=6.5$.
Among them, we randomly select 20 gas particles in each of the simulations and plot their trajectories from $z=15$ to $z=6.5$ (yellow lines).
}

\yueying{
The BH in the low tidal field simulation ($\epsilon = -2\sigma$) has a much higher density surrounding gas at $z=6.5$. 
At a fixed distance range of $r < 3$ $\hckpc$, we end up with significantly more gas particles ($N \sim 1 \times 10^4$) in the left panel than the right panel ($N \sim 5 \times 10^2$).
The spatial distribution of the particles that participate in the BH gas accretion (orange points) shows that the BH neighbouring gas in the low tidal field scenario has a spherical original distribution compared with that in the large tidal field in which the particle participating in the BH accretion originate from a much smaller solid angle.
As further illustrated by the sample trajectories of the right panel, the BH in the large tidal field accretes gas mostly from the $z$ direction and negligibly from the $x$ direction. 
This is because the large tidal field stretches the matter density distribution in the $x$ direction and inhibits the gas accretion onto the central region from the $x$ direction. 
This effect overall delays the BH growth compared with the low tidal field simulation, where gas accretion occurs close to radial trajectories.
}

\begin{figure}
\begin{center}
\includegraphics[width=0.85\columnwidth]{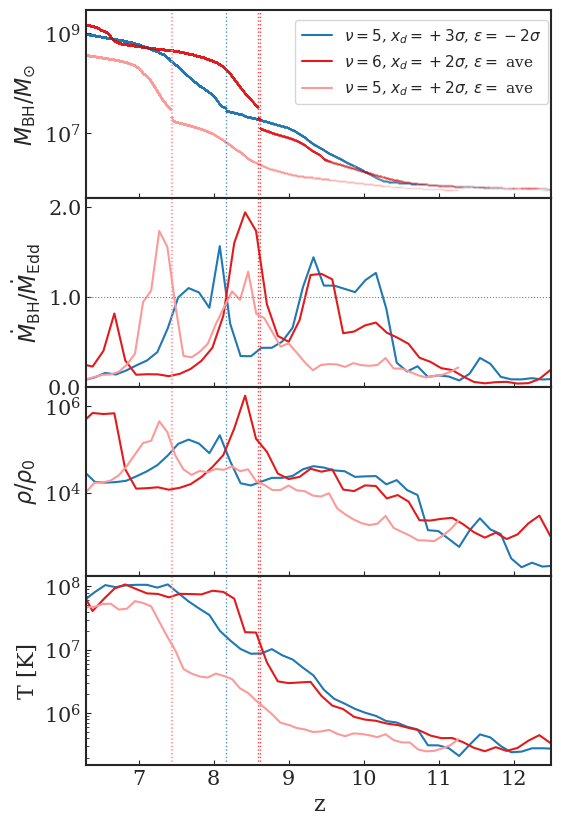}
\end{center}
    \caption{
    Evolution history of the most rapid BH growth in constrained simulations. 
    The blue line represent the result from constrained simulation with $\rm x_d = 5.8$ (+$3\sigma$) and tidal field magnitude $\epsilon=15 \kmsmpc$ ($-2\sigma$). 
    The pink line gives corresponds to constrained peak with compactness $\rm x_d = 5.0$ (+$2\sigma$) and $\epsilon = 34 \kmsmpc$ (ave).
    While the red line gives the result of a constrained density peak with height of $6 \sigma_0$, compactness $\rm x_d = 5.0$ and $\epsilon = 34 \kmsmpc$.
    \textit{Top panel:} Growth of the BH mass as a function of redshift, the gaps in the line correspond to a merging event. 
    \textit{Second panel:} The Eddington ratio $\dot{M}_{\rm BH}/M_{\rm Edd}$ of the BH growth. 
    \textit{Third panel:} The gas density in the vicinity of the BH. The gas density is in the unit of $\rho_0$, the averaged matter density of the universe.
    \textit{Bottom panel:} Averaged gas temperature in the surrounding of BH.
    }
\label{fig:BH-growth-detail}
\end{figure}

\begin{figure}
\begin{center}
\includegraphics[width=1.0\columnwidth]{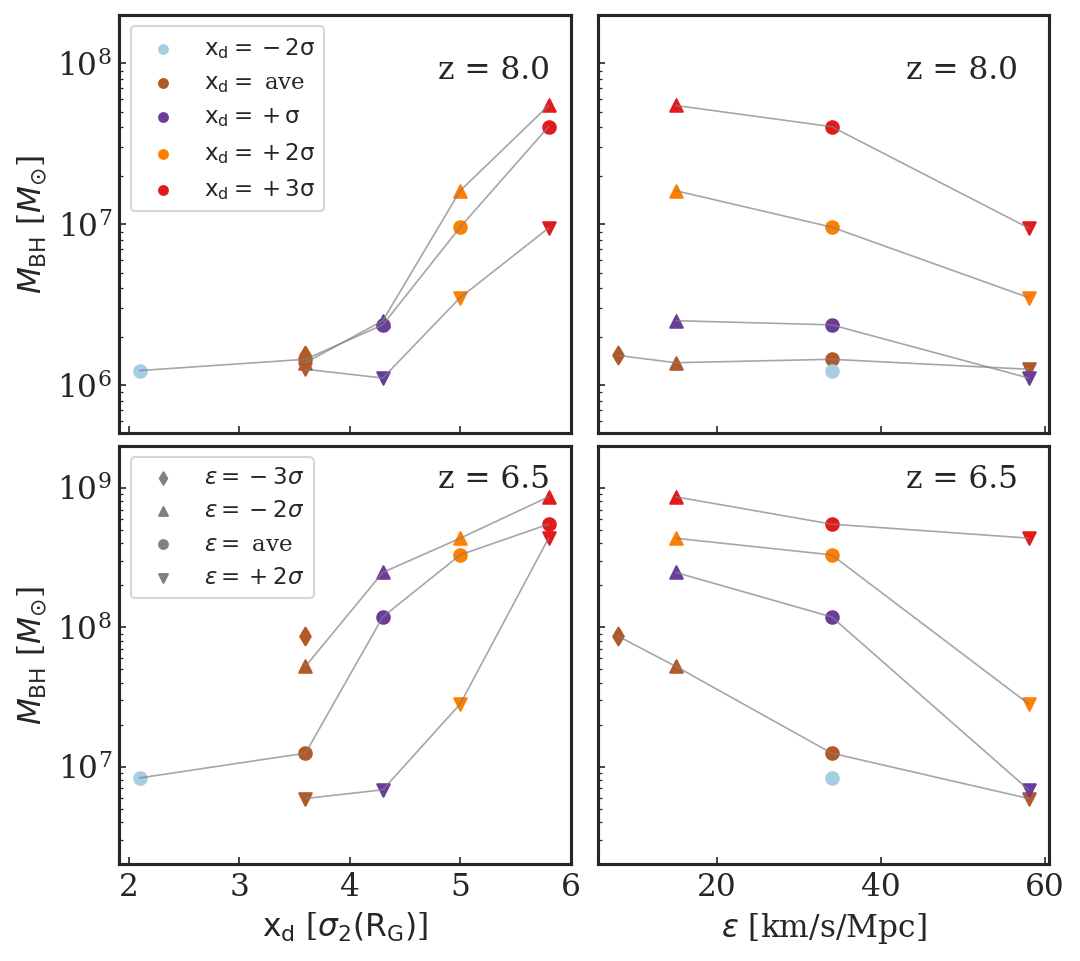}
\end{center}
    \caption{
    A summary of the BH mass in constrained simulations with various peak parameters to show the relation of $M_{\rm BH}$ with the peak compactness $\rm x_d$ (the left column) and tidal field $\epsilon$ (the right column) separately. 
    The upper panels give the BH mass at $z = 8.0$ while the bottom panels give the BH mass at $z = 6.5$.
    The cyan, brown, purple, orange and red colour of the marker represents compactness of the initial density peak $\rm x_d = 2.2$ (-2$\sigma$), 3.6 (ave), 4.3 (+1$\sigma$), 5.0 (+2$\sigma$) and 5.8 (+3$\sigma$) respectively. 
    The different tidal magnitude of the peak is represented by the shape of the marker for  $\epsilon$ = 8 (diamond), 15 (triangle up), 34 (round), 58 (triangle down) in the unit of $\kmsmpc$.
    The grey lines in the left panel link the BHs with the same tidal field magnitude $\epsilon$, and link the BHs with the same peak compactness $\rm x_d$ in the right panels.
    }
    \label{fig:MBH-xd-epsilon}
\end{figure}


\subsubsection{Growth of the most massive BHs}

Given our findings that high compactness and low tidal field of the initial density peak help boosting the BH growth in the early universe, we now intentionally design a set of constrained simulations that give the most rapid BH growth.
In Figure~\ref{fig:BH-growth-detail}, we show the detailed evolutionary history of the most massive BHs in those designed constrained simulations.

As a demonstration of the most rapid and earliest BH growth, the blue line in Figure~\ref{fig:BH-growth-detail} shows the BH growth history with a 5$\sigma_0$ initial density peak and $+3\sigma$ compactness and $-2\sigma$ tidal field. 
We also investigate how the initial height of the density peak (which we have kept fixed until now) may affect the BH growth. 
We run an additional constrained simulation with 6$\sigma_0$ height for the initial density peak with compactness $x_{\rm d} = +2 \sigma$ and mean tidal field, and show the corresponding BH growth history in the red line. 
As a comparison, the pink line plots the result from the constrained simulation with the same compactness and tidal field but with peak height $\nu = 5\sigma_0$ which we have shown earlier in Figure~\ref{fig:BH-growth-xd-epsilon}.

The blue line in Figure~\ref{fig:BH-growth-detail} shows that a 5$\sigma_0$ density peak with high compactness ($+3\sigma$) and low tidal field($-2\sigma$) is able to form a $10^9 M_{\odot}$ BH at $z \sim 6.5$. 
A similar large $M_{\rm BH}$ can be achieved with a more rare ($+6\sigma_0$) density peak, as a higher primordial density peak would form a larger structure earlier and boost the process of BH accretion. 
Both the blue and red line reaches $M_{\rm BH} > 10^9 M_{\odot}$ at $z \sim 6.5$, with the red line giving the most rapid BH growth with the 6$\sigma_0$ peak.

We note that even in the rather extreme constrained simulations (with high compactness and low tidal field in the IC) that boost BH growth in the early phases, the $M_{\rm BH}$ will not keep up a steep growth with time. The curve will eventually flatten as a consequence of self-modulation by AGN feedback.
As discussed in the previous section, when the BH grows larger, it will dump part of its accretion energy into its surroundings, heating and clearing out the nearby fueling gas.
This process, in turn, suppresses the gas accretion onto the BH.
Therefore, we stress that apart from the properties of the initial density peak, the complicated astrophysical feedback processes play a crucial role in the formation mechanism of the first QSOs in the high redshift universe.

\subsubsection{Summary statistics of $M_{\rm BH}$ relation with compactness and tidal field}

\yueying{
To thoroughly explore the effect of peak compactness and tidal field on the early BH growth, we ran 14 constrained simulations of 5$\sigma_0$ initial density peaks in total with different combinations of compactness and tidal field, spanning over peak compactness $x_{\rm d}$ = \{ $-2\sigma$, ave, +1$\sigma$, +2$\sigma$, +3$\sigma$\} and tidal field $\epsilon$ = \{$-3\sigma$, -2$\sigma$, ave +2$\sigma$\}. }
As a summary plot in Figure~\ref{fig:MBH-xd-epsilon}, we show the results for the BH mass at $z=8$ (top panels) and $z=6.5$ (bottom panels) in our constrained simulation sets as a function of the peak compactness $\rm x_d$ (left column) and tidal field magnitude $\epsilon$ (right column).

In the left panels, the grey lines link the BHs with the same tidal field magnitude $\epsilon$, while in the right panels, the grey lines link the BHs with the same peak compactness $\rm x_d$.
The left panel shows a clear trend of increasing $M_{\rm BH}$ with higher peak compactness. 
The relations are even steeper at $z = 8$: BHs residing in peaks with compactness $+1 \sigma$ from the mean are still at the seed mass at $z = 8$, while BHs within high compactness peaks have all exceeded $ 10^7 M_{\odot}$. 

On the other hand, a clear anti-correlation is found between $M_{\rm BH}$ and tidal field magnitude $\epsilon$, as significantly more BH growth occurs in correspondingly smaller tidal field peaks. 
We note that the tidal field, which is determined by the matter distribution on larger scales, affects the BH growth at later times than the peak compactness.
As also shown by Figure~\ref{fig:BH-growth-xd-epsilon}, a compact initial density peak can significantly boost the early BH growth, while the tidal field starts to take effect after the BH grows much beyond the seed mass. 
Therefore, we see that the effect of the tidal field for a highly compact peak is more significant at early times ($z = 8$) since the BH grows earlier in the more compact density peak.
For the less compact peaks, however, the tidal field effect is more apparent at later times, $z = 6.5$.
For BHs with average and $+1 \sigma$ compactness at $z = 6.5$, the BH in a $-2 \sigma$ tidal field is more than 1 order of magnitude more massive than the BHs residing in $+2 \sigma$ tidal field.


\begin{figure*}
\begin{center}
\includegraphics[width=1.9\columnwidth]{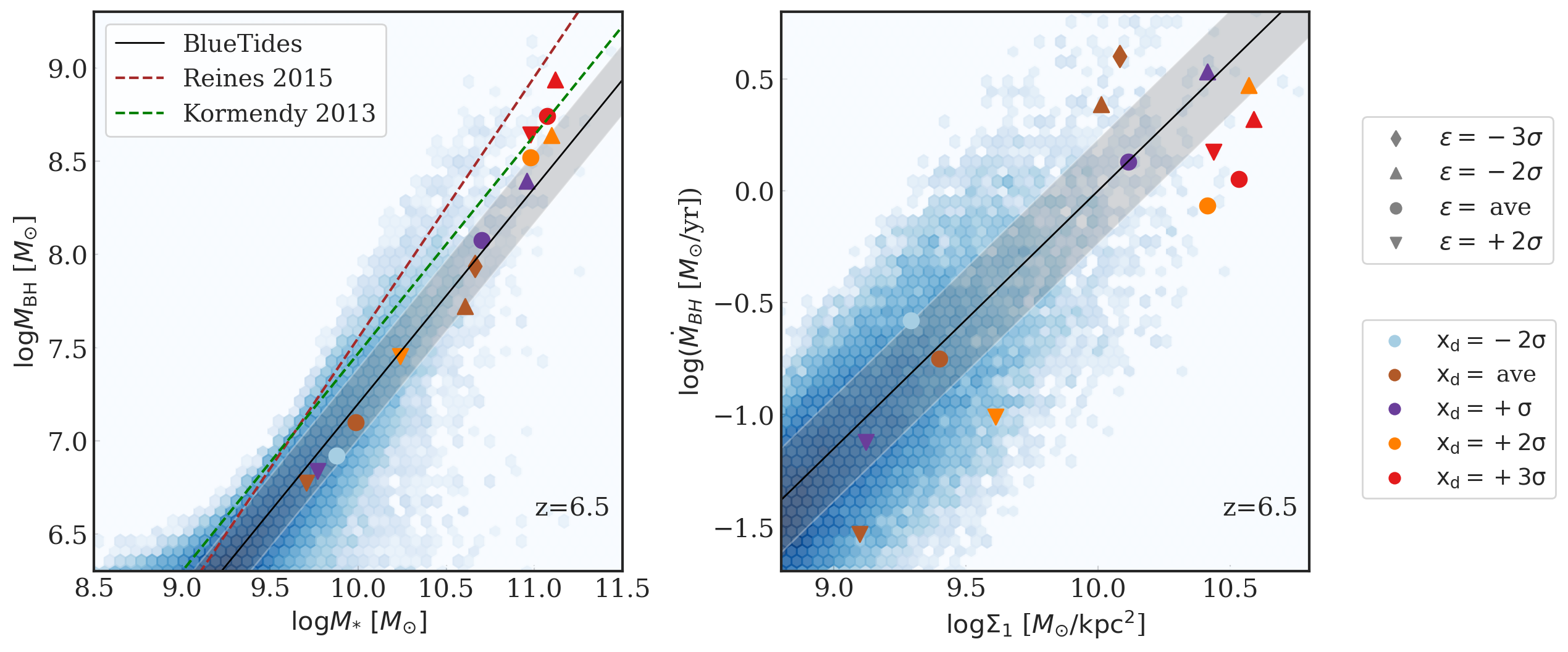}
\end{center}
    \caption{
    \textit{Left panel:} Scaling relation between $M_{\rm BH}$ with respect to the stellar mass in their host galaxy. The colour and shape of the markers represent the peak compactness and tidal field magnitude of the constrained IC, with the same convention as Figure~\ref{fig:MBH-xd-epsilon}.
    The blue shades are from the results of \textsc{BlueTides} simulation.
    The brown dashed line gives the observational result in the local universe from \citet{Reines2016}. 
    The blue dashed line gives the scaling relation from the study of \citet{Kormendy2013}.
    \textit{Right panel:} The relation between the compactness of the host galaxies $\Sigma_1$ with respect to the BH accretion rate. The blue shades are the results of the BH population in \textsc{BlueTides} simulation.
    }
\label{fig:Observation-plane}
\end{figure*}

\begin{figure}
\begin{center}
\includegraphics[width=1\columnwidth]{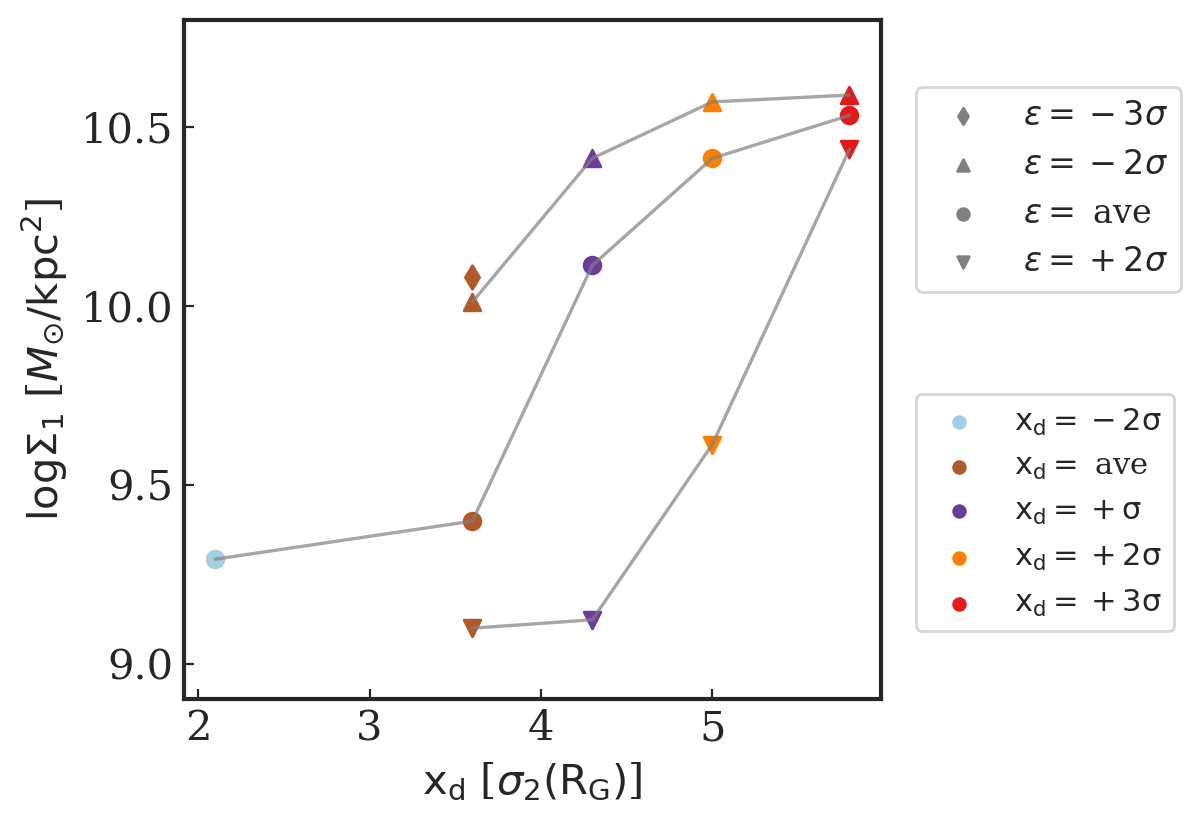}
\end{center}
    \caption{
    The relation between the compactness of the host galaxies at $z=6.5$ with respect to the initial peak parameters of compactness. $y$ axis gives the central surface-mass density within 1 pkpc around central BH $\Sigma_1$ as the representative of the compactness of the host galaxy. The colour and shape of the markers represent the peak compactness and tidal field magnitude of the constrained IC, with the same convention as Figure~\ref{fig:MBH-xd-epsilon}.}
    \label{fig:sigma1-xd-epsilon}
\end{figure}


\section{Implications for Observations}
\label{section5:Obs}

Until now we have only discussed how the BH growth depends on the properties of the peaks in the simulation IC. 
It is well established that SMBHs are connected to the growth of their galaxies \citep[e.g.,][]{Kormendy2013}. 
We will now see how the BHs and their host galaxies compare in our simulations.  
It is also important to examine how the BHs from constrained simulations with 5$\sigma_0$ density peaks compare to the observed BH-galaxy relations.

In the left panel of Figure~\ref{fig:Observation-plane}, we plot the $M_{\rm BH}$ - $M_{*}$ relation for the BHs and their host galaxies in the constrained simulations at $z = 6.5$. 
The different colours and shapes of the markers correspond to the specific peak compactness and tidal field magnitudes of the constrained IC, with the same conventions as for Figure~\ref{fig:MBH-xd-epsilon}.
For comparison, the blue histogram shows the results for the large BH population in the \textsc{BlueTides} simulation.
The black solid line is the linear fitting \yy{between $\log M_{\rm BH}$ and $\log M_{*}$} from \textsc{BlueTides}, with the grey shaded area giving the intrinsic scatter of the fitting.
The brown and \yy{green} dashed lines show fits to the observed $M_{\rm BH}$ - $M_{*}$ relation in the local universe from \cite{Reines2016, Kormendy2013}.

The relation traced by the BHs and galaxies in the constrained simulations is tight and consistent with that extracted from the \textsc{BlueTides} simulation at $z=6.5$. 
This validates the fact that, for the constrained simulations, both BHs and galaxy stellar hosts grow commensurately to the statistical population in \textsc{BlueTides}.
More importantly, the plot clearly shows that there is a large range of BH masses and host stellar masses resulting from a (fixed) given $5 \sigma_0$ initial density peak. This demonstrates that it is not sufficient to have a rare $5 \sigma_0$ initial density peak to lead to a massive BH.
As shown in Figure~\ref{fig:Observation-plane}, $5 \sigma_0$ peaks can populate the lower end of the relation.
As previously emphasized, the constrained runs demonstrate that the different physical properties of the peak can result in a variety of BH masses and associated stellar masses. 
The cluster of points in the highest mass end of the $M_{\rm BH}$ - $M_*$ relation are those corresponding to peaks residing in a low tidal field or with high compactness (or both).
The relatively tight relation traced by the BHs and galaxies in the constrained simulations indicates that the high compactness and low tidal field of the initial density peak lead to both large BH mass and active star formation. 
This in turn leads to stellar components in BH hosts which, at the high mass end, appear consistent with the values inferred from observations.

Interestingly, observations in the local Universe find massive BHs residing in galaxies with notably compact stellar components \citep[e.g.,][]{Walsh2016}.
Several studies of AGNs host galaxies at $z < 3$ also find a positive correlation between BH growth (measured by luminosity) and galaxy compactness which is defined as the surface density of galaxy within the effective radius \citep{Rangel2014, Ni2019}.
These are interesting observations that could provide support to our proposed scenarios for enhanced BH growth in the compact high-density peaks constructed in this study.
Therefore we also examine the relation between the properties of the density peak and the resulting compactness of the stellar component of the host galaxy.
Following the observational measurements, we quantify the galaxy (stellar) compactness by the central surface-(stellar) mass density within 1 pkpc; i.e. we measure
$\Sigma_1 = M_{*}(<1 \rm kpc)/(\pi \times 1 kpc^2)$, where $M_{*}(<1 \rm kpc)$ is the stellar mass enclosed in the central 1 kpc (physical coordinate) around the BH.
\yn{We note that this scale of 1 pkpc is comparable to the half mass radius of the galaxy \citep[see,e.g.][]{Marshall2019}, and therefore is well resolved in the simulation.}

In the right panel of Figure~\ref{fig:Observation-plane}, we plot the relationship between the compactness of the host galaxies, $\Sigma_1$, versus the BH accretion rate in units of $M_{\odot}$/yr from the set of constrained simulations (points) and the \textsc{BlueTides} results for comparison. 
Again, the black solid line shows the linear fitting result from \textsc{BlueTides}, with intrinsic scatter shown in grey.

Overall, $\dot{M}_{\rm BH}$ appears to be correlated with galaxy compactness $\Sigma_1$, though with a rather large scatter (about 0.5 dex). 
The relation supports the observational suggestion that BHs grow more effectively in more compact stellar hosts.
However, we note that there is a rather large scatter in this relation, which is caused by the fact that the BH accretion rate is an instantaneous property and highly variable.
This is shown, for example in Figure~\ref{fig:BH-details-xd-epsilon} and Figure \ref{fig:BH-growth-detail}, where it varies a lot during the growth history corresponding to the interaction with the surrounding gas environment due to AGN feedback. 

A question to investigate is whether the compactness of the stellar host $\Sigma_1$ is indeed related to the compactness of the initial density peak $x_{\rm d}$.
If that is the case, measuring $\Sigma_1$ could help to test our prediction that enhanced BH growth is related to the IC peak compactness.
To show the relation between the stellar compactness and the IC peak parameters, we plot in Figure~\ref{fig:sigma1-xd-epsilon} the results from the constrained simulations at $z = 6.5$, for the compactness of the host galaxies $\Sigma_1$ versus the initial peak compactness parameter $\rm x_d$.
The colour of the data points represents $x_{\rm d}$ and the shape represents tidal field magnitude $\epsilon$ of the IC peak, with the same convention as in Figure~\ref{fig:MBH-xd-epsilon}.

Figure~\ref{fig:sigma1-xd-epsilon} shows a strong positive correlation between IC density compactness $x_{\rm d}$ and the resulting compactness of the host galaxy $\Sigma_1$. 
For the IC peaks with the same tidal field, the one with $x_{\rm d} = +3\sigma$ results in a $\Sigma_1$ about $1 \sim 1.5$ dex higher than the one with mean $x_{\rm d}$.
This indicates that $\rm x_d$ in the initial density field does have an effect on the compactness of the stellar galaxy at later redshift (up to $z=6.5$).
On the other hand, the IC peaks residing in a lower initial tidal field would also lead to a larger $\Sigma_1$, as it helps to form high-density gas clumps around the BH and boosts the star formation. 
This supports our conclusion that the high compactness of the initial density peak and the low tidal field on large scales are favourable to the formation of a high-density gas environment in the halo centre, resulting in a compact galaxy morphology and massive BHs at early times in the universe ($z>6$).

\section{Summary and Conclusions}
\label{section6:conclusion}

In this work, we implement the CR technique introduced by \cite{Hoffman1991,vandeWeygaert1996} to impose constraints on the Gaussian random field of ICs for cosmological simulations.
By building a high density peak in the initial density field, we are able to efficiently form rare massive halos at high redshift $z > 6$ in small cosmological volumes $(20 \hmpc)^3$.
The CR technique also allows us to specify different properties of the initial density peak, as well as sculpt the large scale matter distribution to constrain the characteristics of the gravitational field at the site of the peak.
With the CR implementation, we perform a systematic exploration, with minimal computational effort and at a sufficiently high resolution, of the physical characteristics of the IC density field relevant to the growth of the rare SMBHs in the early universe.

First, to validate our methods, in Section~\ref{section3:show-example1} we apply the CR technique to reproduce the formation of the rare massive halos and BHs found in the \textsc{BlueTides} simulation.
\textsc{BlueTides} is a large volume ($400 \hmpc$ per side) high resolution cosmological hydrodynamical simulation targeting the study of the population of  rare $\sim 10^9 M_{\odot}$ quasars at $z > 6$ \citep{Feng2016}.
We first extract the density peak features in the progenitor region of the quasar hosts from the \textsc{BlueTides} ICs, and impose those peak parameter constraints on a random realization of the initial density field with box size  20 $\hmpc$.
With the constrained ICs, our new simulations have successfully recovered the evolution of the large-scale structure as well as the growth history of the BHs and their halo hosts with reasonable consistency.
More importantly, the demand for computational resources is significantly less, by a factor of $(400 / 20)^3 \sim 8000$.

Previous studies of \textsc{BlueTides} \citep{DiMatteo2017} find that the large density peaks of the first quasars favour some specific physical characteristics, such as a low tidal field environment.
In this work, we run a set of constrained cosmological simulations designed to study the environment and large-scale structures relevant to the growth of the first quasars at $z > 6$.  
In particular, we focus on the initial density peaks with a height of $\nu = 5 \sigma_0$ on the scale of $R_{\rm G} = 1$ $\hmpc$ (corresponding to the hosts of the most massive BHs in  \textsc{BlueTides}) and study the influence of various peak properties on the growth of the first SMBHs.
Such specialized simulations allow us to address the issue of the role of tidal fields in shaping large-scale structures as well as the gas inflows into galaxies that can lead to the fast growth of seed BHs.

We have carried out a series of constrained cosmological simulations with varying peak parameters drawn from the distribution conditioned on peak height $\nu = 5 \sigma_0$.
As a conclusion, we find that the compactness $\rm x_d$ of the initial density peak and the tidal field magnitude $\epsilon$ are two of the most important parameters relevant to the BH growth. 
A more compact initial density field residing in a low tidal field forms a dense, cuspy gas environment in the centre of the halo and therefore induces the most rapid BH growth. 
For example, for $5\sigma$ density peaks with the same tidal field, the more compact one (with $x_{\rm d} = +3 \sigma$) can host BHs two orders of magnitude more massive than the BHs residing in an averagely compact peak at $z=7$.
\yueying{In particular, peak compactness $\rm x_d$ has a larger effect on boosting BH accretion at early epochs. A compact initial density peak leads to an earlier formation of the parent halo and induces a high-density central region fueling the rapid BH growth. On the other side, the tidal field (shaped by the matter distribution on larger scales) starts to take effect at a later stage, after the BH grows much larger than the seed mass. A large tidal field would stretch the matter density distribution, inhibit the gas accretion onto the central region from that direction, and therefore delay the BH growth.}

We also note that, even in the most extreme case of a high density peak with large compactness and low tidal field, the BH growth cannot consistently stay in a high accretion mode. This is because the AGN feedback process will dump significant accretion energy into the  BH surroundings as the BH grows, which will heat and drive out its nearby fueling gas and suppress the accretion process.

Section~\ref{section5:Obs} probes the relation between BH relation and galaxy host in our constrained simulations.
We find a large range of BH masses and host stellar masses resulting from the $5\sigma_0$ initial density peak, with the $M_{\rm BH} - M_*$ relation consistent with the scaling relation predicted from the \textsc{BlueTides} simulation. 
The IC density peaks with large compactness and low tidal field lead to both large BH mass and active star formation. 
Moreover, we find that the host galaxies of BHs in those constrained simulations with higher $\rm x_d$ and lower $\epsilon$ are also more compact in terms of the central stellar mass surface density $\Sigma_1$, indicating that $\rm x_d$ in the initial density field does have consequences for the compactness of the stellar galaxy at later redshift.

\section*{Acknowledgements}
The \textsc{BlueTides} simulation is run on the BlueWaters facility at the National Center for Supercomputing Applications.
Most of the simulations in this work are carried out on the bridges cluster.
Some of the simulations in this work are carried out on the Frontera supercomputing cluster.
The authors also acknowledge the Pittsburgh Supercomputing Center and Texas Advanced Computing Center (TACC) at the University of Texas at Austin for providing HPC resources that have contributed to the research results reported within this paper.
TDM acknowledges funding from NSF ACI-1614853, NSF AST-1616168, NASA ATP 19-ATP19-0084, 80NSSC20K0519
NASA ATP 80NSSC18K101, and NASA ATP NNX17AK56G.

\section*{Data Availability}
Data of the \textsc{BlueTides} simulation is available at http://bluetides.psc.edu;
Data of the constrained simulations generated in this work will be shared on reasonable request to the corresponding author.

\bibliographystyle{mnras}
\bibliography{bib.bib}

\appendix

\section{Effect of the simulation box size}
\label{Appendix A}
\begin{figure}
\begin{center}
\includegraphics[width=0.9\columnwidth]{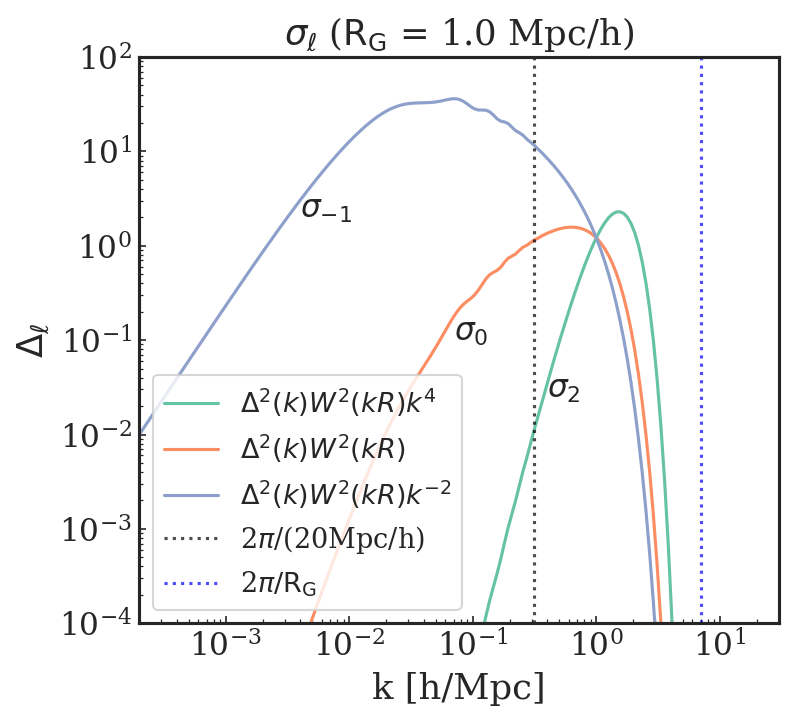}
\end{center}
    \caption{
    Illustration of the scale contribution to the spectral moment $\sigma_{l} (\rm R_G)$ on scale $\rm R_G = 1$ $\hmpc$. The green, orange and blue solid line shows the scale contribution to $\sigma_2$, $\sigma_0$ and $\sigma_{-1}$ respectively. 
    The black dotted line marks out the scale corresponding to the box size of our simulation, corresponding to $k = 2 \pi$/$L_{\mathrm{box}}$, with $L_{\mathrm{box}}$ = 20 $\hmpc$. The blue dotted line gives $k = 2 \pi/\rm R_G$ that corresponds to the cutoff of the solid lines by the smoothing kernel $W(kR)$, since all the variations are calculated by smoothing over the scale of $\rm R_G$.}
    \label{fig:scale-variance}
\end{figure}

Here we briefly discuss the effect of box size for the constrained simulations. 
In this work, we base our studies on the features of the density field on the scale of $\rm R_G = 1$ $\hmpc$.
The variance of some properties of the field could be estimated by the spectral moment given in Eq.~\ref{equation:spectral_moment},
with $\sigma_2$ proportional to the variance of the peak compactness, $\sigma_0$ proportional to the peak height and tidal field strength, while $\sigma_{-1}$ is proportional to the variance of the peculiar velocity.

In Figure~\ref{fig:scale-variance}, we plot the scale contribution to the spectral moment $\sigma_l (\rm R_G)$ on the scale $\rm R_G = 1$ $\hmpc$, with $y$ axis 
\begin{equation}
\begin{split}
    \Delta^2_l &\equiv \Delta^2(k) W^{2}(kR) k^{2l} \\
    &= 1/(2\pi^2) P(k)k^3 \exp(-k^2 R^2) k^{2l}
\end{split}
\end{equation}
showing the variation contributed by each ($\log_{10} k$) bin.
The green, orange and blue solid line shows the scale contribution to $\sigma_2$, $\sigma_0$ and $\sigma_{-1}$ respectively. 
The black dotted line marks out the scale corresponding to our simulation boxsize: $k = 2 \pi /L_{\rm box}$, with $L_{\rm box}$ = 20 $\hmpc$.  
The blue dotted line gives $k = 2 \pi/\rm R_G$ that corresponds to the cutoff of the solid lines by the smoothing kernel $W(kR)$, since all the variations are calculated by smoothing over the scale of $\rm R_G$.

Note that Figure~\ref{fig:scale-variance} shows $y$ axis in log scale.
We can see that $\sigma_2$ and $\sigma_0$ are mostly contributed by the scale within the box size. 
However, for $\sigma_{-1}$ which corresponds to the peculiar velocity, the variance is mostly contributed by large scale ($k = 0.01 \sim 0.1 \invhmpc$). 
In other words, the peculiar velocity of the density field is mostly generated by the variation of matter distribution on a scale larger than our box size.
Though it is always possible to constrain a large peculiar velocity on an arbitrary position of the simulation box, this is achieved by enforcing a large asymmetry in local matter distribution wrapped by the periodic boundary of the simulation box.
The matter distribution would be significantly different if we change the box size while imposing the same velocity constraint. 
Therefore we caveat that it is improper to constrain a large peculiar velocity up to ($\sigma_{-1} \sim 300 \kms$) in our small box.

One limitation of our CR implementation is that the super-sampling variance is missing \citep{Li2014}.
In our simulation, we (implicitly) assume that the overdensity (i.e. the DC mode) and also the tidal field of the simulation box to be zero, and therefore can not model the matter evolution response to the modes larger than our box size. 

The DC mode can be incorporated by the so-called separate universe (SU) technique that absorbs the overdensity of the simulation volume into a modified cosmology.
\citep[see, e.g.,][for more details]{Sirko2005,Gnedin2011,Wagner2015,Li2014,Li2018}.
Recently, the SU technique has also been extended to incorporate the large scale tidal field \citep{Akitsu2020}.
However, we also note that the long modes on the scales larger than our box size 20 $\hmpc$ are hardly relevant to the BH growth at this high redshift regime of $z>6$. 
Future work can be carried out to combine the CR and SU in simulations, and further investigate the long mode effects on the BH evolution in more detail.

\section{Illustration of peak parameters}
\label{Appendix B}

\begin{figure*}
\begin{center}
\includegraphics[width=1.9\columnwidth]{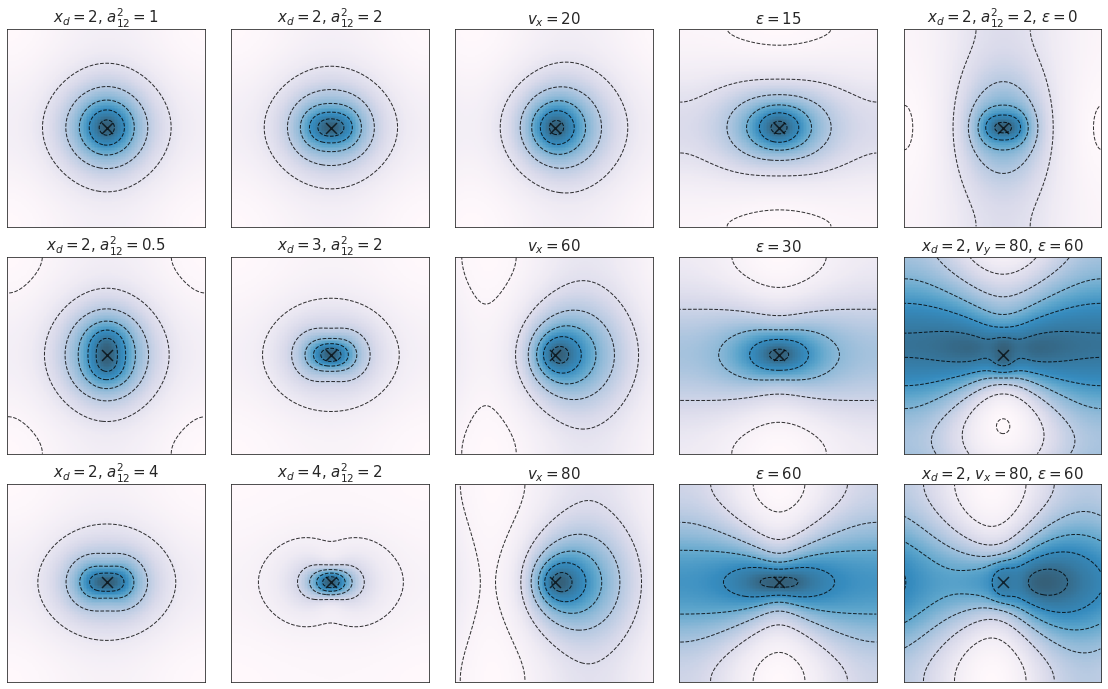}
\includegraphics[width=1.9\columnwidth]{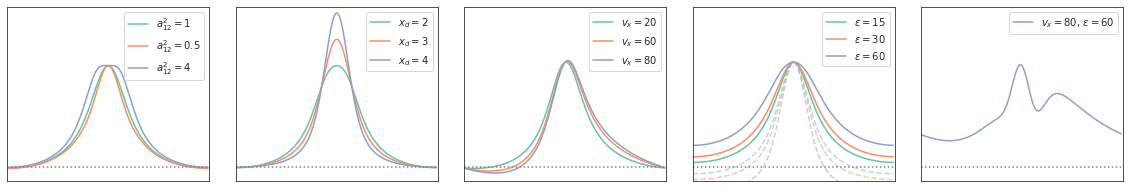}
\end{center}
    \caption{Illustration of the effects of the variation of different peak parameters. In each panel, we show the ensemble mean field of a density peak with height $\nu_c = 3 \sigma_0(\rm R_G)$ subject to a subset of the other peak constraints. 
    All the constraints are imposed at the centre of the box (as marked by the black cross), with a scale of $\rm R_G = 1$ $\hmpc$. The box is 20 $\hmpc$ per side projected onto the $xy$ plane. 
    The black dashed lines in each panel show the isodensity contours of the density peak projected onto the 2D plane.
    The coloured solid lines in the bottom panels of each column show the $\delta_x$ profile along $x$ axis crossing the peak, with green, red and blue colours corresponding to the top, middle and bottom panels above. The grey dotted line shows where $\delta_x = 0$.
    \textbf{The first and second columns}: the Ensemble mean field $\bar{f}(\bf x)$ constructed with \{$\hat{H}_i (\bf k)$, | i=1,$5 \sim 10$ \}, illustrating the second order derivatives of the peak. The first column fixes the compactness $\rm x_d = 2 \sigma_2 (\rm R_G)$, and variate $a^2_{12} = 1$, 0.5, 4 in each panel to illustrate ellipticity, where $a_{12}$ is the axial ratio between the $x$ and $y$ coordinate.
    The second column fixes $a^2_{12} = 2$ and variate $\rm x_d$ = 2, 3, 4 to illustrate the effect of the peak compactness.
    \textbf{The third column}: The $\bar{f}(\bf x)$ field constructed with $\hat{H}_1 (\bf k)$ and the peculiar velocity field in the $x$ direction $\hat{H}_{11} (\bf k)$, to illustrate the matter distribution of a density peak with peculiar velocity  $v_x$ = 20, 60, 80 $\kms$ respectively. \textbf{The fourth column}: The $\bar{f}(\bf x)$ field constructed with $\hat{H}_1 (\bf k)$ and $\hat{H}_{14} (\bf k)$ $\sim \hat{H}_{18} (\bf k)$ to illustrate the effect of the tidal field magnitude with $\epsilon$ = 15, 30, 60 $\kmsmpc$ respectively. The shear angle is $\omega = 1.5 \pi$. We make the tidal field elongated in the $x$ direction and compressed in the $y$ direction.
    \textbf{The fifth column}: The $\bar{f}(\bf x)$ field constructed with the full 18 peak constraints to illustrate the effect of a combination of the peak parameters. 
    See text for more details.} 
    \label{fig:params_separate}
\end{figure*}

In this section, we give a more detailed illustration of the effects of various peak parameters.
Given a density peak in the smoothed density field convolved by Gaussian kernel of width $\rm R_G$, we have the 15 parameter sets $\{$ $\nu$,$x_d$,$a_{12}$,$a_{13}$,$\alpha_1$,$\beta_1$,$\gamma_1$, $v_x$,$v_y$,$v_z$,$\epsilon$, $\omega$,$\alpha_2$,$\beta_2$,$\gamma_2$ $\}$ to characterize the features of the peak. 

To illustrate the effects of these peak parameters on the density field, we plot in Figure~\ref{fig:params_separate} the ensemble mean field $\bar{f}(\bf x)$ of a density peak with height $\nu_c = 3 \sigma_0(\rm R_G)$ subject to a subset of the peak constraints to show the effect of the peak parameters separately.
The $\bar{f}(\bf x)$ field is constructed via Eq.~\ref{equation:Ensemble_mean} as a superposition of the $\xi_i(\bf x)$ fields built from the corresponding $\hat{H}_i (\bf k)$. 
The constraint position $\mathbf{r}_\mathrm{pk}$ is at the centre of the box, as marked by the black cross in each panel.  
The box is 20 $\hmpc$ per side projected onto the $xy$ plane. 
All the constraints are imposed on a scale of $\rm R_G = 1$ $\hmpc$.

The first two columns in Figure~\ref{fig:params_separate} illustrate the effect of varying the second order derivatives of the density peak. 
The $\bar{f}(\bf x)$ field in each panel is built with a weighted superposition of \{$\xi_i$, | i=1,$5 \sim 10$ \} (c.f., Eq.~\ref{equation:Ensemble_mean}). 
The first column sets the compactness of the peak to be $\rm x_d = 2 \sigma_2 (\rm R_G)$, $a^2_{13} = 1$, and the three Euler angles to be 0 so that $a_{12}$ corresponds to the axial ratio of the mass ellipsoid along the $x$ and $y$ coordinates. 
We plot the $\bar{f}(\bf x)$ field with $a^2_{12} = 1$, 0.5, 4 in the top, middle and bottom panels to illustrate the ellipticity of the peak. 
In the second column, we fix $a^2_{12} = 2$, set $a^2_{13}$ and the Euler angles to be the same as in the first column, and assign variate $\rm x_d$ = 2, 3, 4$\sigma_2(\rm R_G)$ in the three panels to illustrate the effect of the peak compactness.
As shown by the density profile in the bottom, the density peak with higher compactness has a more concentrated matter distribution. 

The third column of Figure~\ref{fig:params_separate} shows the $\bar{f}(\bf x)$ field constructed with $\hat{H}_1 (\bf k)$ and the peculiar velocity field in the $x$ direction $\hat{H}_{11} (\bf k)$, to illustrate the matter distribution of a density peaks with peculiar velocities $v_x$ = 20, 60, 80 $\kms$ in the positive $x$ direction respectively. 
As shown by the density profile, the overdensity to the right of the peak (positive $x$ direction) attracts the matter from the left, and therefore induces the peculiar velocity of the density peak in the positive $x$ direction.

The fourth column shows the $\bar{f}(\bf x)$ field constructed using $\hat{H}_1 (\bf k)$ and $\hat{H}_{14} (\bf k)$ $\sim \hat{H}_{18} (\bf k)$ to illustrate the effect of the tidal field magnitude with $\epsilon$ = 15, 30, 60 $\kmsmpc$ in the top, middle and bottom panels respectively.
Here we set the shear angle $\omega = 1.5 \pi$ (as illustrated by the third panel of Figure~\ref{fig:contour3D}), and set the corresponding Euler angles of the tidal field so that the density peak is elongated in the $x$ direction and compressed in the $y$ direction. The bottom panel of the fourth column plot shows using dashed lines the $\delta_x$ profile along the $y$ axis so that we can see that the density peak residing in a larger tidal field is stretched in the $x$ direction and squeezed in $y$ direction.

The fifth column gives the $\bar{f}(\bf x)$ field constructed using the full 18 peak constraints to illustrate the effect of a combination of the peak parameters. 
The top panel sets $\rm x_d = 2\sigma_2(\rm R_G)$, $a^2_{12}=2$ and $\epsilon=0$. Naturally, as shown from the top panel of the second column, $a^2_{12}=2$ has an extended matter distribution along the $x$ axis which would imply a non-zero tidal field with elongation in the $x$ direction. 
However, since we add the further constraint that the tidal field magnitude of the peak $\epsilon=0$, the construction of the $\bar{f}(\bf x)$ field will modulate the larger scale structure correspondingly to compensate for the tidal force contributed by the immediate surroundings of the peak. Therefore we see that the isodensity contour forms a vertical ellipsoid on large scales.
The second panel show the $\bar{f}(\bf x)$ field with $\rm x_d = 2\sigma_2(\rm R_G)$, $a^2_{12}=1$,$v_y = 80 \kms$ and $\epsilon=60$ $\kmsmpc$, so that the matter distribution is elongated in the $x$ direction, and compressed in the $y$ direction, with an overdensity in positive $y$.
The third panel shows the $\bar{f}(\bf x)$ field $\rm x_d = 2\sigma_2(\rm R_G)$, $a^2_{12}=1$,$v_x = 80 \kms$, with $\epsilon=60$ $\kmsmpc$, and with the corresponding $\delta_x$ profile along the $x$ axis shown in the bottom panel.

\section{Constraint kernels}
\label{Appendix C}

Here, we plot the full 18 $\xi_i (\bf x)$ fields constructed via $\hat{H}_1 (\bf k)$ $\sim \hat{H}_{18} (\bf k)$ (c.f. Eq.~\ref{equation:xi_i}) projected on $xy$ plane. 
See Section~\ref{section2:Method} for details.

\begin{figure*}
\begin{center}
\includegraphics[width=1.9\columnwidth]{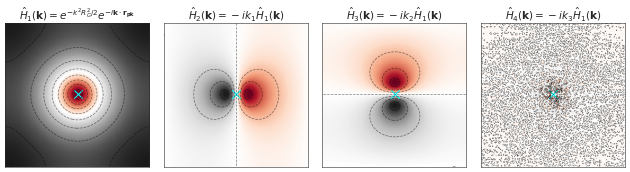}
\includegraphics[width=1.4\columnwidth]{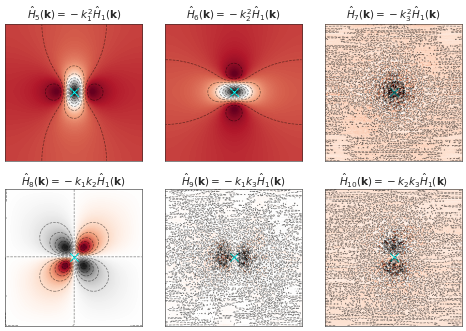}
\includegraphics[width=1.9\columnwidth]{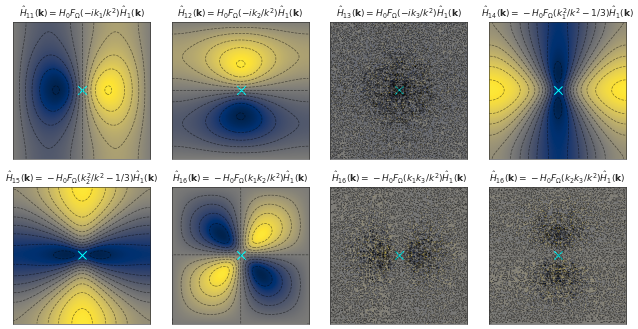}
\end{center}
    \caption{Illustration of the full 18 $\xi_i (\bf x)$ fields shown in the xy plane, with $\rm R_G = 1$ $\hmpc$ and peak position located at the centre of the 20 $\hmpc$ box. The first 10 kernels shape the immediate surrounding of the density peak. While the last 8 kernels put constraints on the gravitational field that can sculpt the larger range of the matter distribution.}
    \label{fig:H18_full}
\end{figure*}

\end{document}